\documentclass[11pt,twoside]{article}
\usepackage{axodraw}
\usepackage{graphicx}
\usepackage{colordvi}
\usepackage{color}
\usepackage{epsf}
\usepackage{latexsym}
\usepackage{epsfig}
\begin{document}
\begin{center}
{\Large \textbf{SUPERSYMMETRY AND LHC}}

\vspace{10mm}

\large A.~V.~Gladyshev and  D.~I.~Kazakov

\normalsize \vspace{5mm} \textit{Bogoliubov Laboratory of Theoretical Physics, Joint
Institute for Nuclear Research, Dubna}

\vspace{1mm}
 \textit{Institute for Theoretical and Experimental Physics, Moscow}
\end{center}
\vspace{5mm}
\begin{abstract}
The motivation for introduction of supersymmetry in high energy physics as well as a
possibility for supersymmetry discovery at LHC (Large Hadronic Collider) are discussed.
The main notions of the Minimal Supersymmetric Standard Model (MSSM) are introduced.
Different regions of parameter space are analyzed and their phenomenological properties
are compared. Discovery potential of LHC for the planned luminosity is shown for
different channels. The properties of SUSY Higgs bosons are studied and perspectives of
their observation at LHC are briefly outlined.
\end{abstract}
\tableofcontents \vglue 0.1cm \noindent {\bf 10 \ References \hfill 27}
%
\section{Introduction}
Supersymmetry or symmetry between bosons (particles with integer
spin) and fermions (particles with half-integer spin) has been
introduced in theoretical papers nearly 30 years ago~\cite{super}.
Since that time there appeared thousands of papers, all quantum
field theory models were supersymmetrized, new mathematical tools
were derived that allow one to work with anticommuting variables.
The reason for this remarkable activity is the unique mathematical
nature of supersymmetric theories, possible solution of various
problems of the Standard Model of fundamental interactions within
its supersymmetric extentions as well as the opening perspective of
unification of all interactions in the framework of a single
theory~\cite{susy}.

Supersymmetry today is the main candidate for a unified theory
beyond the Standard Model. Search for various manifestations of
supersymmetry in Nature is one of the main tasks of numerous
experiments at colliders and in non-accelerator experiments of the
last decade. Unfortunately, the result is negative so far. There are
no any direct indications on existence of supersymmetry in particle
physics  though existing supersymmetric models satisfy all
theoretical and experimental requirements. Remarkably that the scale
of supersymmetry breaking, or as it is often said the scale of new
physics, is about 1~TeV what is 10 times bigger than the electroweak
symmetry breaking scale at which the LEP accelerator was adjusted.
And it is this energy scale that the LHC accelerator will explore.
It is assumed that at LHC the TeV energy range will be examined in
detail, the Higgs boson will be found and supersymmetry will be
discovered.

Supersymmetry is the challenge for the world physics community which
was accepted with construction of LHC.  Thus, high energy physics
approaches the crucial moment when low energy  supersymmetry will be
either discovered or abandoned. One has to be ready for such
circumstances  and clearly realize which signatures of supersymmetry
one can expect and how to extract them from that sea of data which
will be obtained at the two main detectors of LHC: ATLAS and CMS.
\section{Motivation of Supersymmetry}

Recall what are the main arguments in favour of supersymmetric extension of the Standard
Model of fundamental interactions. Though these arguments are not new their
attractiveness does not weaken with time. They include
\begin{itemize}
\item
unification with gravity. This is perhaps the main argument in favour of supersymmetry
within the unification paradigm. The point is that SUSY algebra being a generalization of
Poincar\'e algebra  links together representations with different spins. The key relation
is given by the anticommutator
$$\{Q_\alpha, \bar{Q}_{\dot \alpha}\}=2\sigma_{\alpha,\dot
\alpha}^\mu P_\mu .$$
 Taking infinitesimal transformations
$\delta_\epsilon = \epsilon^\alpha Q_\alpha, \ \bar{\delta}_{\bar \epsilon} =
\bar{Q}_{\dot \alpha}{\bar \epsilon}^{\dot \alpha},$ one gets
 \begin{equation}
\{\delta_\epsilon,\bar{\delta}_{\bar \epsilon} \}
 =2(\epsilon \sigma^\mu \bar \epsilon )P_\mu ,
 \label{com}
 \end{equation}
where $\epsilon$ is a transformation parameter. Choosing $\epsilon$ to be local, i.e. a
function of a space-time point $\epsilon = \epsilon(x)$, one finds from eq.(\ref{com})
that an anticommutator of two SUSY transformations is a local coordinate translation. And
a theory which is invariant under the general coordinate transformation is General
Relativity. Thus, making SUSY local, one obtains General Relativity, or a theory of
gravity, or supergravity~\cite{sugra}.

\item
unification of gauge couplings. According to {\em hypothesis} of Grand Unification gauge
symmetry increases with energy. All known interactions are the branches of a single
interaction associated with a simple gauge group which includes the group of the SM as a
subgroup. Unification (or splitting) occurs at very high energy ($10^{15} \div
10^{16}$~GeV).

To reach this goal one has to examine how the coupling change with energy. This is
described by the renormalization group equations. In the leading order of perturbation
theory  solutions take a simple form:
\begin{equation}
\frac{1}{\alpha_i(Q^2)} = \frac{1}{\alpha_i(\mu^2)}- b_i
\log(\frac{Q^2}{\mu^2}),
\label{alphasol}
\end{equation}
where index $i$ refers to the gauge groups $SU(3) \times SU(2) \times U(1)$, and for the
SM one has $b_i=(41/10,-19/6,-7)$. Result is demonstrated in Fig.~\ref{fig:unif}, where
the evolution of the inverse couplings is shown as functions of log of energy. In the
left part of Fig.~\ref{fig:unif} one can see that in the SM unification of the gauge
couplings is impossible. In the supersymmetric case the slopes of RG curves are changed,
for the minimal supersymmetric extension of the SM one has $b_i=(33/5,1,-3)$. It happens
that in supersymmetric model one can achieve perfect unification as it is shown in the
right part of Fig.~\ref{fig:unif}. Fitting the curves one can get the scale of SUSY
breaking $M_{SUSY}\sim 1$~TýÂ~\cite{ABF}.
\begin{figure}[htb]\begin{center}
\includegraphics[width=0.85\textwidth]{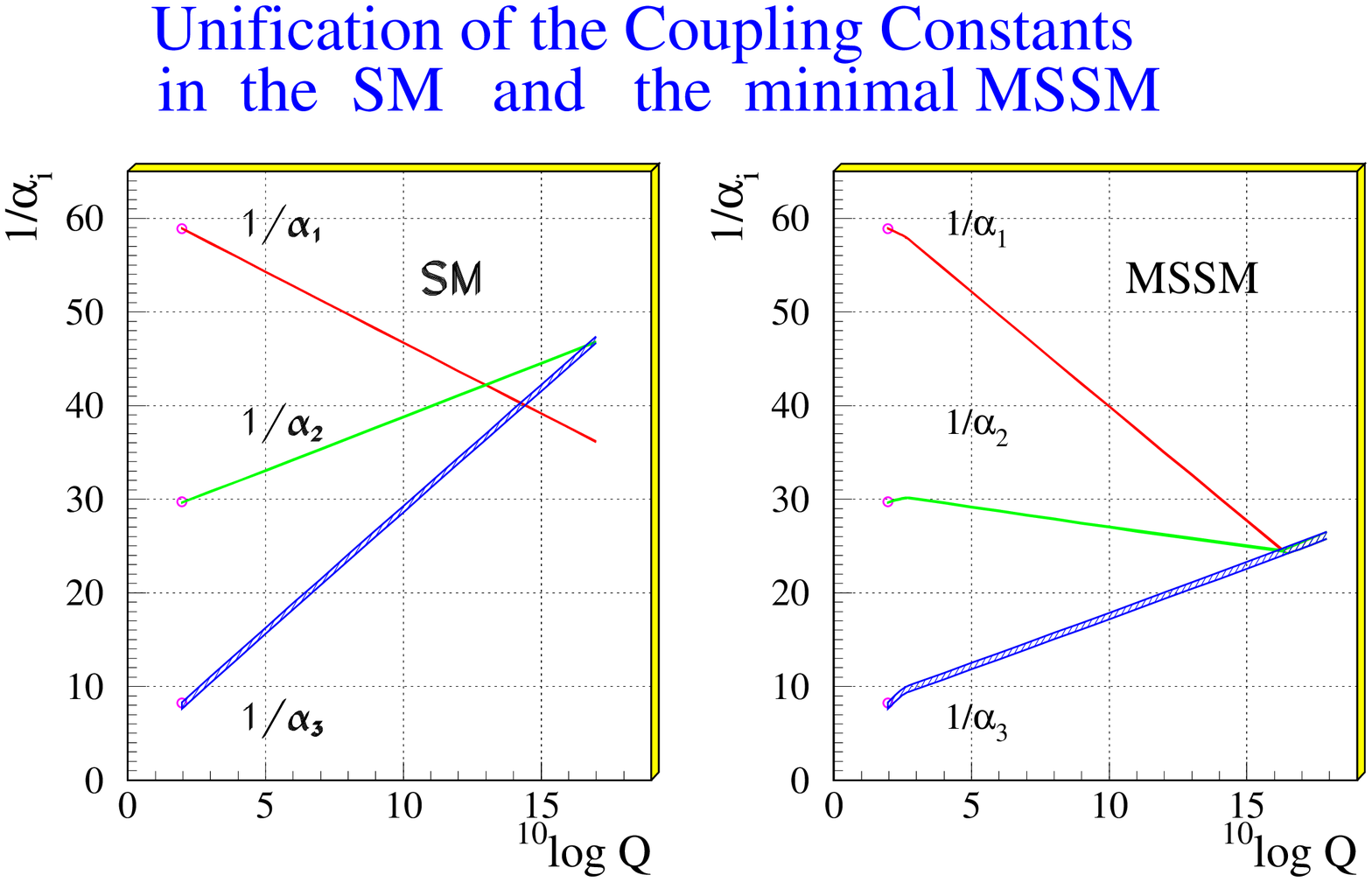}
 \caption{Evolution of the inverse gauge couplings in the SM (left) and in the MSSM (right).}
  \label{fig:unif}\end{center}
\end{figure}
\item
solution of the hierarchy problem. The appearance of two different
scales in Grand Unified theories, namely $M_Z \ll M_{GUT}$, leads to
the very serious problem which is called the {\em hierarchy
problem}. First, this is the very existence of the hierarchy.
Second, is the preservation of a given hierarchy in presence of the
radiative corrections. These corrections, proportional to the mass
of a heavy particle, inevitably destroy the hierarchy unless they
are cancelled. The only way to get this cancellation of quadratic
mass terms (also known as the cancellation of quadratic divergences)
is supersymmetry. Moreover, supersymmetry automatically cancels all
quadratic corrections in all orders of perturbation theory due to
the contributions of superpartners of the ordinary particles. The
contributions of the boson loops are cancelled by those of fermions
due to additional factor $(-1)$ coming from Fermi statistic. This
cancellation is true up to the SUSY breaking scale, $M_{SUSY}$,
since
\begin{equation}
 \sum_{bosons} m^2 - \sum_{fermions} m^2= M_{SUSY}^2,
\end{equation}
which  should not be very large ($\leq$ 1 TeV) to make the fine-tuning natural. Indeed,
let us take the Higgs boson mass. Requiring for consistency of perturbation theory that
the radiative corrections to the Higgs  boson mass do not  exceed the mass itself gives
 \begin{equation}
\delta M_h^2 \sim g^2 M^2_{SUSY} \sim M_h^2.\label{del}
 \end{equation}
 Thus, we
again get the same rough estimate of $M_{SUSY}\sim M_Z/g\sim 10^3$~GeV as from the gauge
coupling unification above. Two requirements match together.

The origin of the hierarchy is  the other part of the problem. We show below how SUSY can
explain this part as well.
\item
radiative electroweak symmetry breaking. The "running" of the Higgs
masses leads to the phenomenon known as {\em radiative electroweak
symmetry breaking}. Indeed, one can see from
Fig.~\ref{fig:mass-evol} that the mass parameters from the Higgs
potential $m_1^2$ and $m_2^2$ (or one of them) decrease while
running from the GUT scale to the scale $M_Z$ may even change the
sign. As a result for some value of the momentum $Q^2$ the potential
may acquire a nontrivial minimum. This triggers spontaneous breaking
of $SU(2)$ symmetry. The vacuum expectations of the Higgs fields
acquire nonzero values and provide masses to quarks, leptons and
$SU(2)$ gauge bosons, and additional masses to their superpartners.
\begin{figure}[htb]
\begin{center}
\includegraphics[width=0.45\textwidth]{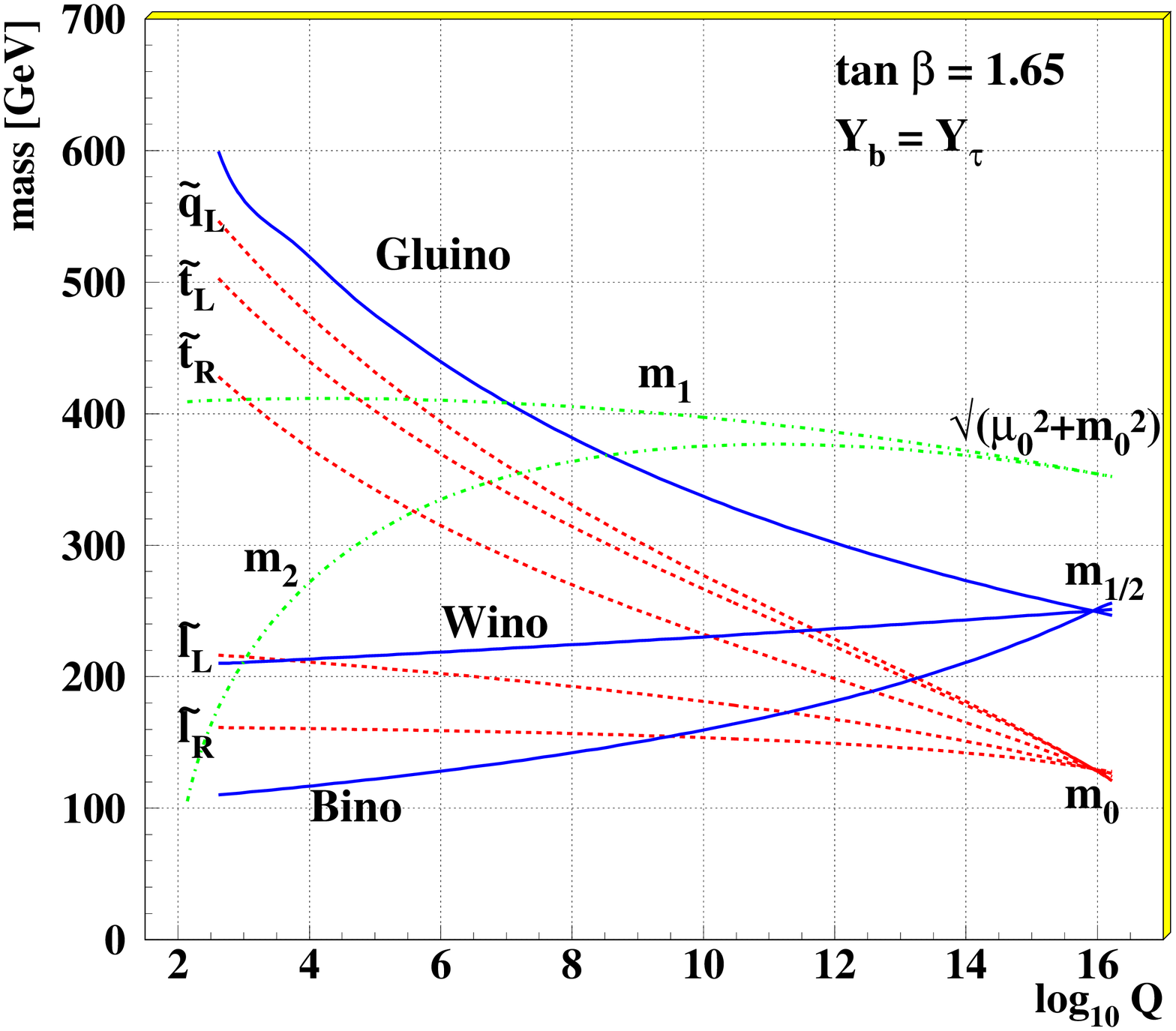}
\includegraphics[width=0.45\textwidth]{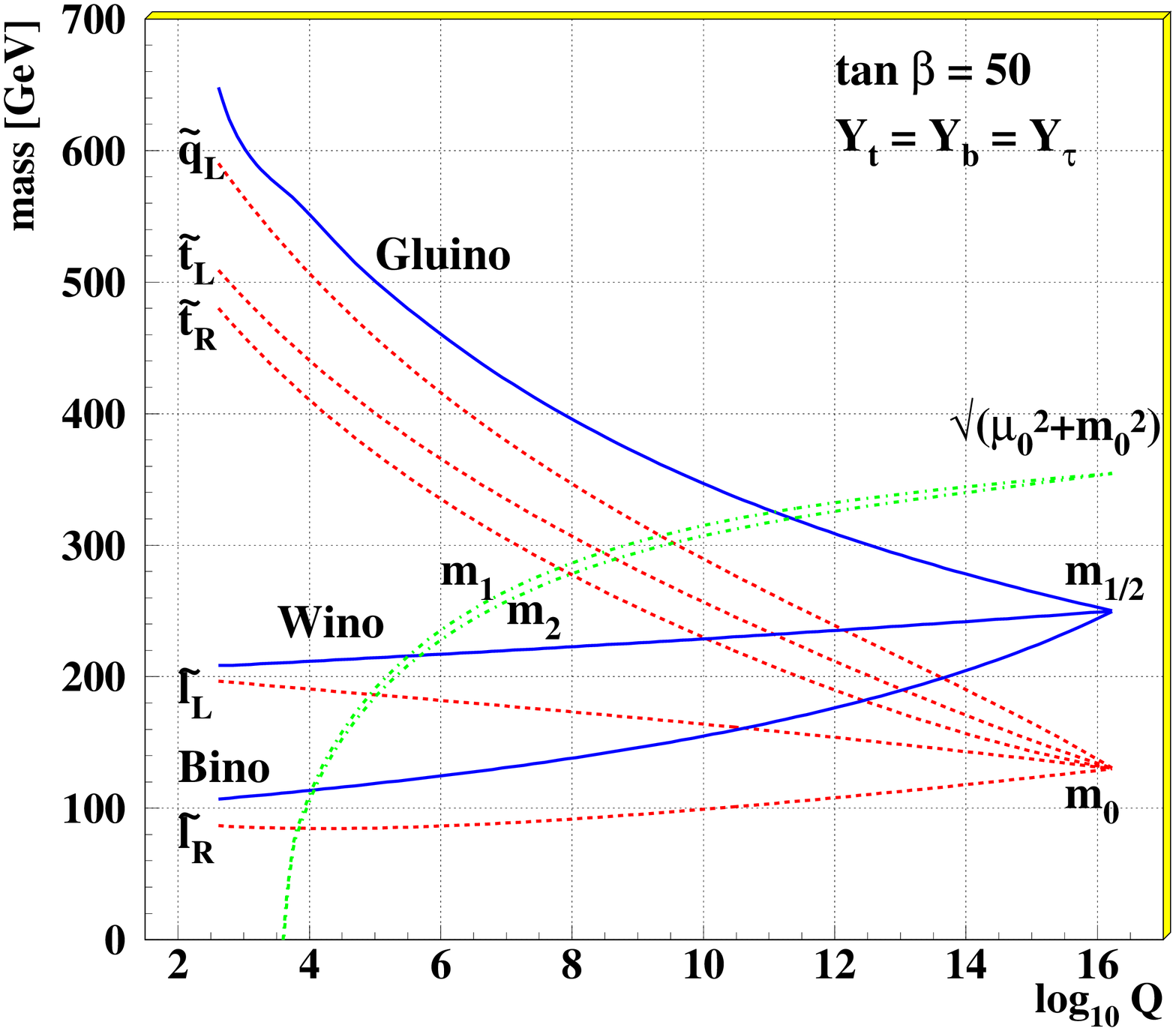}
\end{center}
\caption{Typical evolution of SUSY masses and soft SUSY breaking parameters
$m_1^2=m^2_{H_1}+\mu^2$ and $m_2^2=m^2_{H_2}+\mu^2$ for small (left) and large (right)
values of $\tan\beta$~\cite{running}.} \label{fig:mass-evol}
\end{figure}
Thus, the breaking of the electroweak symmetry is not introduced by
a brute force as in the SM, but appears naturally from the radiative
corrections. In this way one also obtains the explanation of why the
two scales are so much different. Due to the logarithmic running of
the parameters, one needs  a long "running time" to get $m_2^2$ (or
both $m_1^2$ and $m_2^2$) to be negative when starting from a
positive value of the order of $M_{SUSY}\sim 10^2 \div 10^3$ GeV at
the GUT scale.
\item
Dark matter in the Universe. The visible (or shining) matter is not
the only matter in the Universe. Considerable amount of matter is
the so-called \textit{Dark matter}. Direct indication on the
existence of the Dark matter are rotation curves of spiral galaxies.
To explain these curves one usually assumes the existence of a
galactic halo consisting of non-shining matter which takes part in
gravitational interaction. According to recent data~\cite{WMAP}, the
matter content of the Universe is the following:
$$\Omega_{total}=1.02 \pm 0.02$$
$$ \Omega_{vacuum}=0.73 \pm 0.04, \ \ \Omega_{matter}=0.23 \pm 0.04,
\ \ \Omega_{barion}=0.044 \pm 0.004 \%,$$ i.~e. Dark matter makes up
a considerable part exceeding the visible barionic matter by the
order of magnitude.

There are two possible kinds of nonbarionic Dark matter: hot DM,
consisting of light relativistic particles, and cold DM, consisting
of weakly interacting massive particles (WIMPs). Hot DM might
consist of neutrino, however, this is problematic from the point of
view of large structure formation in the Universe. Besides,
neutrinos are too light to produce enough DM. As for the cold DM, in
the SM there are no appropriate particles. At the same time,
supersymmetry provides an excellent candidate for this role, namely,
neutralino, the lightest superparticle. It is stable, so that the
relic neutralinos might survive in the Universe since the Big Bang.
\end{itemize}

\section{MSSM: the field content and Lagrangian}

Despite complexity of the mathematical structure of supersymmetric
gauge theories, any supersymmetric extension of the Standard Model
has some general simple features which do not depend on a particular
model. This is, first of all, the doubling of particles: each
particle of the SM, quark or lepton or the gauge boson like photon,
gluon or intermediate weak boson, has a partner with the same
quantum numbers but with the spin differing by $1/2$. These
particles are called superpartners. Note that the usual particles of
the SM can not be partners of each other since in the SM one has no
particles with the same quantum numbers and different spin.

The field content of the Minimal Supersymmetric Standard Model
(MSSM) is shown in Table~\ref{tab:mssm} (hereafter the tilde over
the symbol of a particle denotes a superpartner of a usual
particle).
\begin{table}[h]
\caption{\bf Particle Content of the MSSM} \label{tab:mssm} \vglue 0.6cm
\renewcommand{\tabcolsep}{0.03cm}
\begin{tabular}{lllccc}
Superfield & \ \ \ \ \ \ \ Bosons & \ \ \ \ \ \ \ Fermions &
$SU_c(3)$& $SU_L(2)$ & $U_Y(1)$ \\ \hline \hline Gauge  &&&&& \\
${\bf G^a}$   & gluon \ \ \ \ \ \ \ \ \ \ \ \ \ \ \  $g^a$ & gluino$ \ \ \ \ \ \ \ \ \ \
\ \ \tilde{g}^a$ & 8 & 1 & 0 \\ ${\bf V^k}$ & Weak \ \ \ $W^k$ \ $(W^\pm, Z)$ & wino,
zino \ $\tilde{w}^k$ \ $(\tilde{w}^\pm, \tilde{z})$ & 1 & 3& 0 \\ ${\bf V'}$   &
Hypercharge  \ \ \ $B\ (\gamma)$ & bino \ \ \ \ \ \ \ \ \ \ \ $\tilde{b}(\tilde{\gamma
})$ & 1 & 1& 0 \\ \hline Matter &&&&
\\ $\begin{array}{c} {\bf L_i} \\ {\bf E_i}\end{array}$ & sleptons
\ $\left\{
\begin{array}{l} \tilde{L}_i=(\tilde{\nu},\tilde e)_L \\ \tilde{E}_i =\tilde
e_R \end{array} \right. $ & leptons \ $\left\{ \begin{array}{l} L_i=(\nu,e)_L
\\ E_i=e_R \end{array} \right.$ & $\begin{array}{l} 1 \\ 1 \end{array} $  &
$\begin{array}{l} 2 \\ 1 \end{array} $ & $\begin{array}{r} -1 \\ 2
\end{array} $ \\ $\begin{array}{c} {\bf Q_i} \\ {\bf U_i} \\ {\bf D_i}
\end{array}$ & squarks \ $\left\{ \begin{array}{l}
\tilde{Q}_i=(\tilde{u},\tilde d)_L \\ \tilde{U}_i =\tilde u_R \\
\tilde{D}_i =\tilde d_R\end{array}  \right. $ & quarks \ $\left\{
\begin{array}{l} Q_i=(u,d)_L \\ U_i=u_R^c \\ D_i=d_R^c \end{array}
\right.$ & $\begin{array}{l} 3
\\ 3^* \\ 3^* \end{array} $  & $\begin{array}{l} 2 \\ 1 \\ 1 \end{array} $ &
$\begin{array}{r} 1/3 \\ -4/3 \\ 2/3 \end{array} $ \\ \hline Higgs &&&& \\
$\begin{array}{c} {\bf H_1} \\ {\bf H_2}\end{array}$ & Higgses \ $\left\{
\begin{array}{l} H_1 \\ H_2 \end{array}  \right. $ & higgsinos \ $\left\{
 \begin{array}{l} \tilde{H}_1 \\ \tilde{H}_2 \end{array} \right.$ &
$\begin{array}{l} 1 \\ 1 \end{array} $  & $\begin{array}{l} 2 \\ 2
\end{array} $ &
$\begin{array}{r} -1 \\ 1
\end{array} $
 \\ \hline \hline
\end{tabular}
\end{table}
\vglue .5cm
The labels L or R for squarks or sleptons do not mean that they are left or
right handed. Being  spin zero particles they have no handedness. This is used to mark
that they are superpartners
of left or right handed quarks and leptons.

The presence  of the additional Higgs boson is a generic property of
the supersymmetric theory. In the MSSM there are two doublets of
scalar fields with quantum numbers  $(1,2,-1)$ and $(1,2,1)$.

Table~\ref{tab:mssm} does not contain gravitational fields. In the
simplest version of supergravity one has to add to the set of the
MSSM fields  the pair of graviton and gravitino, the particle with
spin 3/2.

The other important feature is the breaking of supersymmetry. If
this does not happen the superpartners would be degenerate in masses
with the ordinary particles, which is not observed. Due to
supersymmetry breaking this degeneracy disappears and superpartners
acquire large masses that explains their non-observation at the
moment. However, the trace of supersymmetry should remain in
relations between the amplitudes of various processes (with
participation of the usual particles and superpartners) and in
contributions of superpartners to the radiative corrections below
the threshold. The concrete predictions depend on the details of
supersymmetry breaking mechanism which is not known yet.

The Lagrangian of the MSSM consists of two parts; the first part is SUSY generalization
of the Standard Model, while the second one represents the SUSY breaking as mentioned
above. The supersymmetric part of the Lagrangian consists of the gauge invariant kinetic
terms corresponding to the $SU(3)$, $SU(2)$ è $U(1)$ gauge groups depending on 3 gauge
couplings as in the Standard Model and of the superpotential. Usually the superpotential
is chosen in the form repeating that of the Yukawa interaction in the SM
\begin{equation}
W = \epsilon_{ij}(y^U_{ab}Q_a^j U^c_bH_2^i +
y^D_{ab}Q_a^jD^c_bH_1^i
       +  y^L_{ab}L_a^jE^c_bH_1^i + \mu H_1^iH_2^j), \label{R}
\end{equation}
where $i,j=1,2,3$ are the $SU(2)$ and $a,b=1,2,3$ are the generation
indices; colour indices are suppressed, $y^{U,D,L}$ are the Yukawa
couplings. This part of the Lagrangian almost exactly repeats that
of the SM except that the fields are now the superfields rather than
the ordinary fields of the SM. The only difference is the last term
which describes the Higgs mixing. It is absent in the SM since there
is only one Higgs field there.

In principle the superpotential can contain other interactions:
\begin{equation}
W_{NR} =  \epsilon_{ij}(\lambda^L_{abd}L_a^i L_b^jE_d^c +
\lambda^{L\prime}_{abd}L_a^iQ_b^jD_d^c +\mu'_aL^i_aH_2^j)
+\lambda^B_{abd}U_a^cD_b^cD_d^c. \label{NR}
\end{equation}
These terms are absent in the SM. The reason is very simple: one can
not replace the superfields in eq.(\ref{NR}) by the ordinary fields
like in eq.(\ref{R}) because of the Lorentz invariance. These terms
have a different property, they violate either lepton (the first
three terms in eq.(\ref{NR})) or baryon number (the last term).
Since both effects are not observed in Nature, these terms must be
suppressed or excluded. One can avoid such terms by introducing a
special symmetry called the $R$-parity~\cite{R} defined by
 \begin{equation}
R=(-1)^{3(B-L)+2S} \label{par}
 \end{equation}
where $B$ --- is the baryon number, $L$ --- is the lepton number,
and $S$ --- is the spin of the particle. Conservation of the
$R$-parity has important phenomenological consequences:
superparticles are created in pairs and  the lightest supersymmetric
particle (LSP) is absolutely stable. This makes the LSP  an
excellent candidate for the Dark matter particle that is one of
attractive features of supersymmetric extension of the SM.

Since non of the fields of the MSSM can have nonzero vacuum
expectation value, needed for the SUSY breaking, without violating
the gauge invariance, it is assumed that spontaneous breaking of
supersymmetry takes place with the help of some other fields. The
most popular scenario of getting low-energy broken supersymmetry is
the so-called {\it hidden sector} scenario~\cite{hidden}. According
to it there are two sectors: the usual matter belongs to the
"visible" sector, while the the second, "hidden" sector, contains
the fields that break supersymmetry.  These two sectors interact
with each other by exchanging some fields called {\it messengers}.
They transport supersymmetry breaking from the hidden sector to the
visible one. The messengers may be various fields: gravitons, gauge
bosons, etc.

SUSY breaking terms of the Lagrangian are often called the {\it soft
terms} since they are the operators of dimension 2 and 3. They
contain a vast number of free parameters which spoils the predictive
power of the model. To reduce their number, we adopt the so-called
universality hypothesis, i.e., we assume the universality or
equality of various soft parameters at the high energy scale,
namely, we put  all the spin 0 particle masses to be equal to the
universal value $m_0$, all the spin 1/2 particle (gaugino) masses to
be equal to $m_{1/2}$ and all the cubic and quadratic terms repeat
the structure of the Yukawa superpotential (\ref{R}). This is an
additional requirement motivated by the supergravity mechanism of
SUSY breaking. Universality is not a necessary requirement and one
may consider nonuniversal soft terms as well. However, it will not
change the qualitative picture presented below; so for simplicity,
in what follows we consider the universal boundary conditions. In
this case, the soft terms take the form
\begin{eqnarray}
-{\cal L}_{Breaking} & = & m_0^2\sum_{i}^{}|\varphi_i|^2 +\left( \frac 12
m_{1/2}\sum_{\alpha}^{} \tilde \lambda_\alpha\tilde \lambda_\alpha\right.
\label{soft2}\\
&  & \left.\hspace*{-2.2cm}  +\ A[y^U_{ab}\tilde Q_a \tilde U^c_bH_2+y^D_{ab}\tilde Q_a
\tilde D^c_bH_1+ y^L_{ab}\tilde L_a\tilde E^c_bH_1] +  B[\mu H_1H_2] + h.c.\right),
\nonumber
\end{eqnarray}
where $\varphi$ denote the scalar fields of squarks, sleptons and
the Higgs bosons, $\lambda$ --- are the gauginos, the spinor
superpartners of the gauge fields, $A$ and $B$
--- are the new parameters of dimension of a mass.

Thus, the Minimal Supersymmetric Standard Model has the following free parameters: i)
three gauge couplings $\alpha_i$; ii) three matrices of the Yukawa couplings $y^i_{ab}$,
where $i = L,U,D$; iii) the Higgs field mixing parameter  $\mu $; iv) the soft
supersymmetry breaking parameters. Compared to the SM there is an additional Higgs mixing
parameter, but the Higgs self-coupling, which is arbitrary in the SM,  is fixed by
supersymmetry. The main uncertainty comes from the unknown soft terms.

With the universality hypothesis one is left with the following set of 5 free parameters
defining the mass scales
 $$ \mu, \ m_0, \ m_{1/2}, \ A \ \mbox{and}\
 B \leftrightarrow \tan\beta = \frac{v_2}{v_1}. $$
Instead of parameter $B$ one usually uses the parameter $\tan\beta
\equiv v_2/v_1$ equal to the ratio of v.e.v's of the Higgs fields.
Choosing the values of these free parameters one can predict the
mass spectrum of superpartners and the cross-sections for their
production.

\section{Supersymmetry breaking: the parameter space}

To reduce arbitrariness in the choice of the MSSM parameters and to make more definite
predictions one usually imposes several constraints which also serve as a consistency
checks of the model. As it happens, in the MSSM one can simultaneously fulfil several
such constraints:
\begin{itemize}
\item
Gauge coupling constant unification. This is one of the most
restrictive constraints. It fixes the scale of SUSY breaking of the
order of 1 TeV.
\item
$M_Z$ from electroweak symmetry breaking;\\
Radiative EW symmetry breaking  defines the mass of the $Z$-boson
\begin{equation}
\label{defmz} \frac{M_Z^2}{2}=\frac{m_1^2-m_2^2\tan^2\beta}{\tan^2\beta-1}=-\mu^2+
\frac{m_{H1}^2-m_{H2}^2\tan^2\beta}{\tan^2\beta-1}.
\end{equation} This condition determines the value of $\mu^2$ for given
values of $m_0$ and $m_{1/2}$. The sign of $\mu$ remains undefined,
it can be fixed from the other constraints.
\item
 Yukawa coupling constant
unification. The masses of top, bottom and $\tau$ can be obtained from the low energy
values of the running Yukawa couplings via
 \begin{equation} m_t=y_t\
v\sin\beta, \ \ m_b=y_b\ v\cos\beta, \ \ m_\tau=y_\tau \ v\cos\beta . \label{yuk}
\end{equation}
They can be translated to the pole masses taking into account the
radiative corrections, which restricts possible solutions in the
Grand Unified Theories.
\item
Precision measurement of decay rates. Radiative corrections due to
superpartners may essentially influence the decay rates under the
threshold. The typical example is the branching ratio $BR(b\to s
\gamma)$ which has been measured by BaBar, CLEO and BELLE
collaborations~\cite{bsg} and yields the world average of $BR(b\to s
\gamma)=(3.43\pm0.36)\cdot 10^{-4}$. The Standard Model contribution
to this process gives slightly lower result, thus leaving a window
for SUSY. This requirement imposes severe restrictions on the
parameter space, especially for the case of large $\tan\beta$.
\item
Anomalous magnetic moment of muon. Recent measurement of the
anomalous magnetic moment indicates small deviation from the SM of
the order of 2 $\sigma$\cite{am}: $\Delta
a_\mu=a_\mu^{exp}-a_\mu^{theor}=(27\pm10)\cdot 10^{-10}$ . The
deficiency may be easily filled with SUSY contribution, which is
proportional to $\mu$ and $\tan\beta$. This requires positive sign
of $\mu$ and kills a half of the parameter space of the
MSSM~\cite{Anom}.
\item
Experimental lower limits on SUSY masses. SUSY particles have not
been found so far and from the searches at LEP one knows  the lower
limit on the charged lepton and chargino masses of about one half of
the centre of mass energy~\cite{LEPSUSY}. The lower limit on the
neutralino masses is smaller. There exist also limits on squark and
gluino masses from the hadron colliders~\cite{TEVSUSY}.  These
limits restrict the  minimal values for the SUSY mass parameters.
\item
Dark Matter constraint. Recent precise astrophysical data restrict the amount of the Dark
Matter in the Universe to  $23\pm 4$~\%. Assuming that the Hubble constant is $h_0
\approx 0.7$ one finds that the contribution of each relict particle $\chi$ has to obey
the constraint $\Omega_\chi h^2_0 \sim 0.12 \pm 0.02$ and serve as a very severe bound on
SUSY parameters~\cite{relictst} that leaves a very narrow band of allowed region in
parameter space.
\end{itemize}

Requirement of simultaneous fulfilment of these constraints defines
the allowed regions of parameter space. However, not all of the
above mentioned parameters are equally important. Besides, some of
them are practically not free, since they are severely constrained.
For example, as we already mentioned, the Higgs mixing parameter
$\mu$ is related to  $m_0$, $m_{1/2}$ and $Z$-boson mass. The triple
coupling $A$ in many cases is inessential and its value at the GUT
scale is often chosen to be $A_0=0$. The requirement of Yukawa
coupling unification restricts the value of $\tan\beta$. There are
two possible scenaria: scenario with small $\tan\beta$ ($\tan\beta
\approx 1 \div 3$) and scenario with large $\tan\beta$ ($\tan\beta
\approx 30 \div 70$)~\cite{tan}. These scenaria are rather different
from phenomenological point of view since the allowed regions of
parameter space are different. Unfortunately, the recent LEP data
practically exclude the small $\tan\beta$ scenario since the mass of
the lightest Higgs boson happens to be below the experimental limit.
Besides, the astrophysical data are also in favour of large
$\tan\beta$. Thus, from the set of free parameters of the model we
basically have two independent ones: $m_0$ and $m_{1/2}$. It is very
useful therefore to use the $(m_0,m_{1/2})$ plane to present the
theoretical and experimental constraints that we discussed above.
Since the scale of supersymmetry breaking is of the order of 1~TeV,
the masses of superpartners should be in the same region, which
defines the range of parameters $m_0$ and $m_{1/2}$.

\begin{figure}[htb]
\begin{center}
\includegraphics[width=0.40\textwidth]{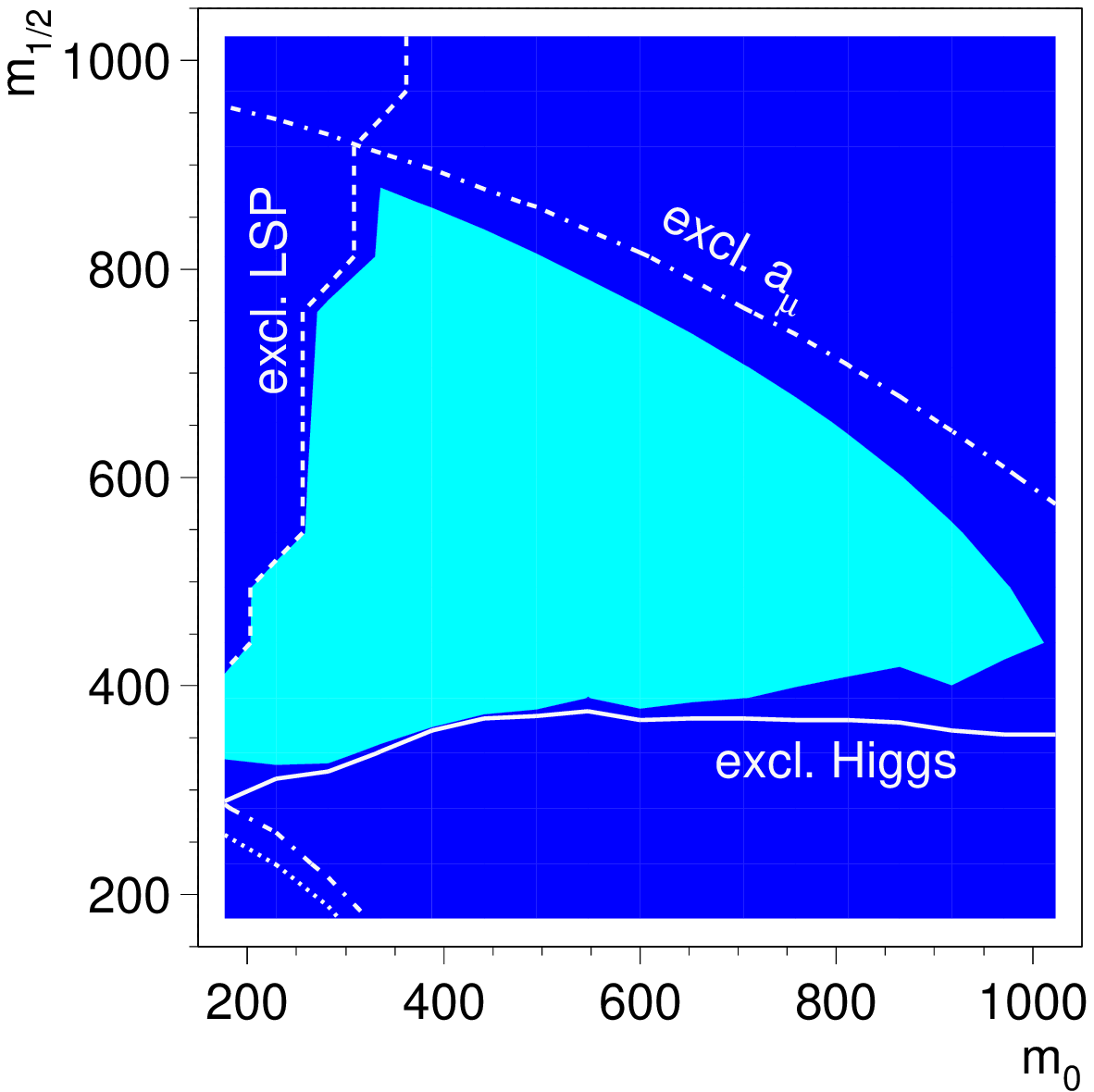}
\includegraphics[width=0.40\textwidth]{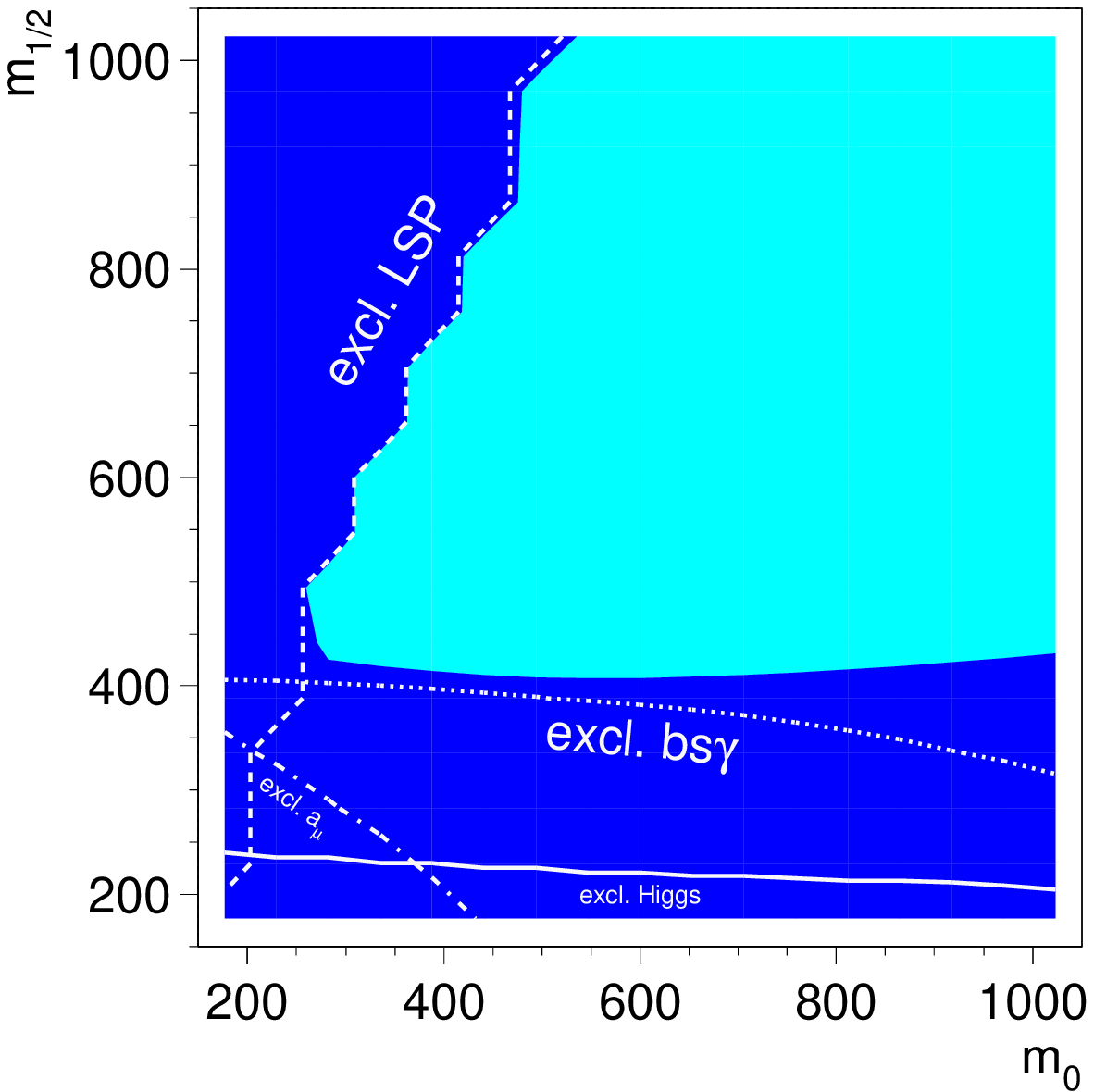}
\end{center}
\caption{The allowed regions of parameter space for the large  $\tan\beta$ scenaria for
different constraints (left -- $\tan\beta=35$, right -- $\tan\beta=50$).}
\label{fig:region}
\end{figure}

We consider further how each of the above mentioned constraints cuts
out the allowed regions in the parameter space in the plane
$(m_0,m_{1/2})$. In Fig.~\ref{fig:region} these regions are shown
for two fixed values of  $\tan\beta=35$ and $\tan\beta=50$ without
account of astrophysical data yet for the values  of $m_0$ and
$m_{1/2}$ in the interval from 200 to 1000~GeV.
\begin{figure}[htb]
\begin{center}
\includegraphics[width=0.40\textwidth]{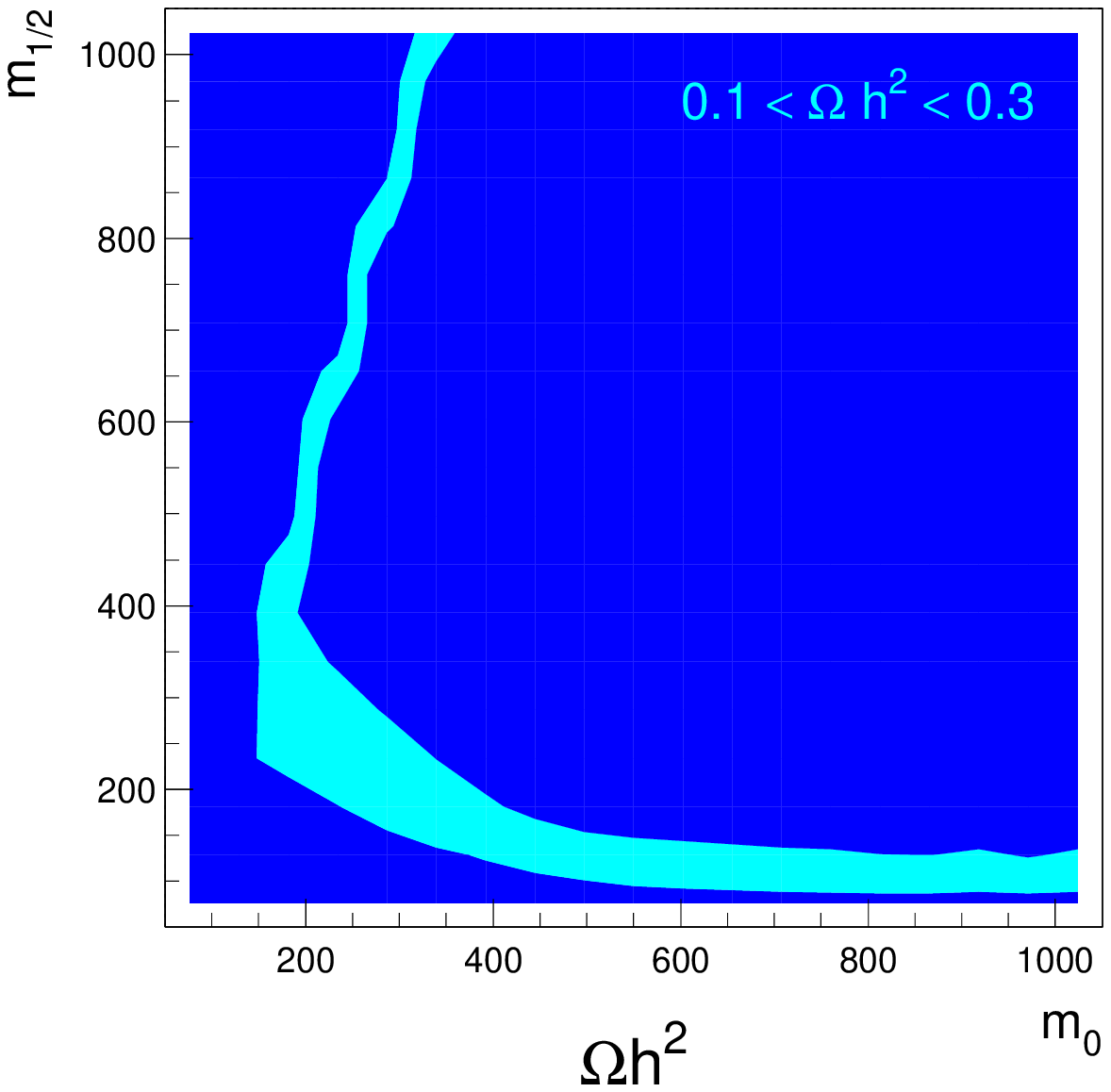}
\includegraphics[width=0.40\textwidth]{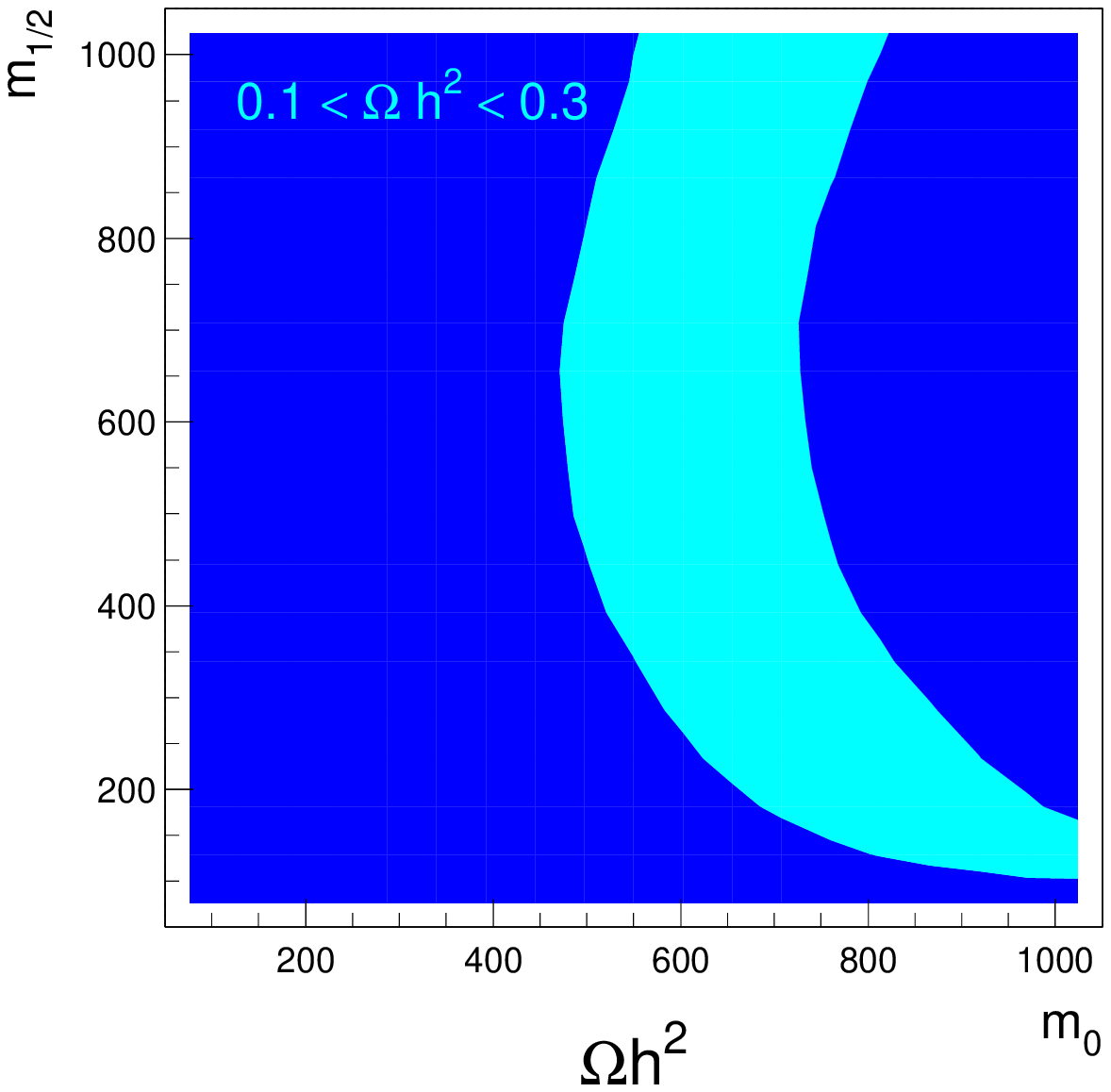}
\end{center}
\caption{Constraints on the parameter space from the requirement of right amount of the
Dark matter (left - $\tan\beta=35$, right - $\tan\beta=50$).} \label{fig:astro}
\end{figure}
%
\begin{figure}[hb]
\begin{center}
\includegraphics[width=0.5\textwidth]{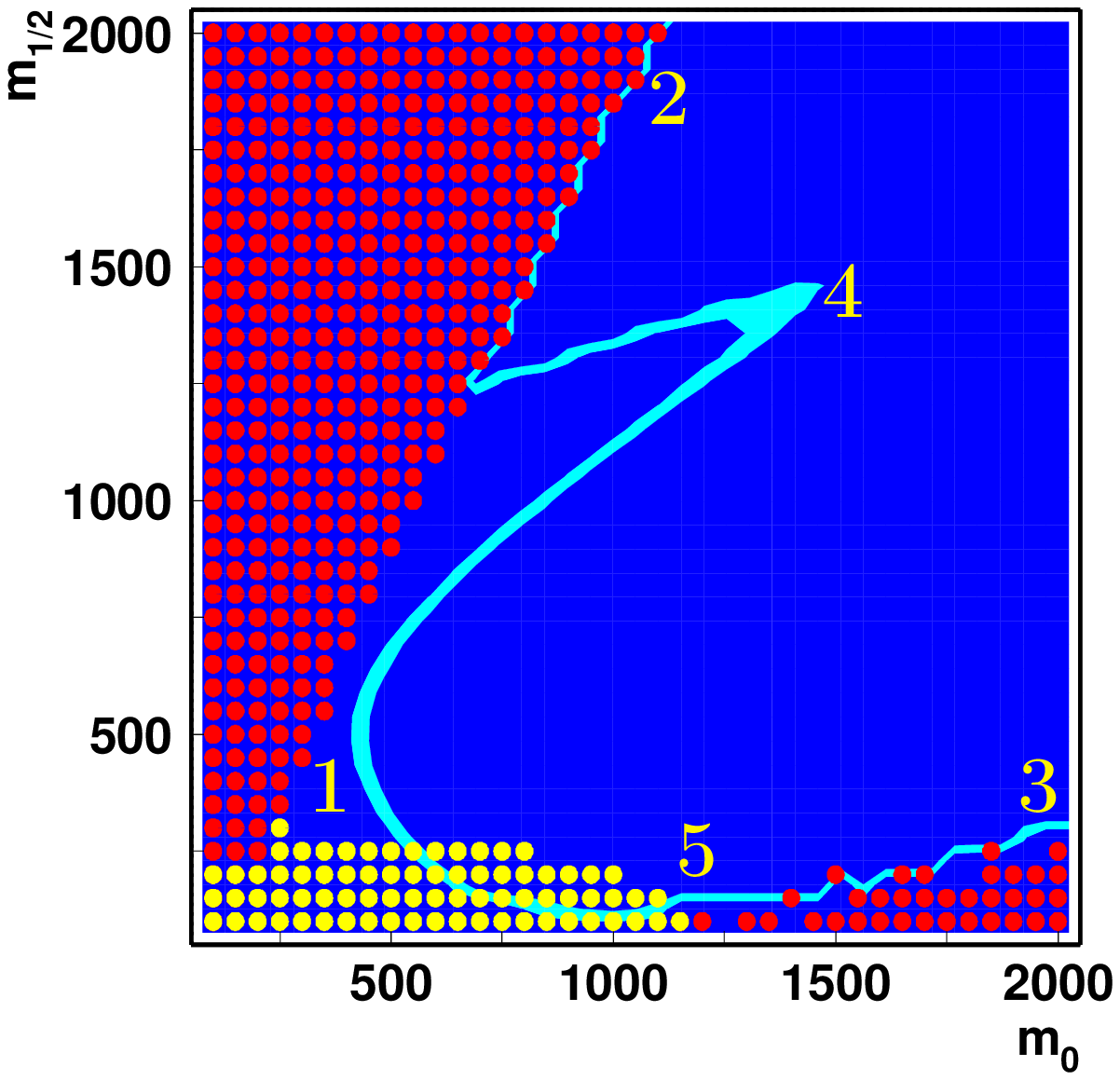}
\caption{The light (lbue) band is the region allowed by the WMAP data for $\tan\beta=51$,
$\mu>0$ and $A_0=0.5m_0$. The excluded regions where the LSP is stau (red up left
corner), where the radiative electroweak symmetry breaking mechanism does not work (red
low right corner), and where the Higgs boson is too light (yellow lower left corner) are
shown with dots. The numbers denote: 1 --- the main annihilation region, 2
--- the co-annihilation region , 3 --- the focus point region, 4 --- the funnel region, 5 ---
the EGRET region.} \label{fig:WMAP}\end{center}
\end{figure}
We start with the radiative electroweak symmetry breaking. It
happens so that for very large $m_0$ and small  $m_{1/2}$  the mass
parameters of the scalar potential, which start from $m_0$, do not
have enough "time" to run to negative values when the conditions of
non-trivial minima of the potential are satisfied. Therefore the
left low corner of $(m_0,m_{1/2})$ plane is usually excluded by this
requirement.

Similar thing happens with the constraint related to the
non-observation of the Higgs boson, which forbids the region of
small values of $m_{1/2}$ practically independently of $m_0$ for
fixed $\tan\beta$. As for the $\tan\beta$ dependence it is the
following: the smaller the value of $\tan\beta$ the large the
excluded region. This fact basically excludes the small $\tan\beta$
scenario as incompatible with the LEP lower limit on the Higgs boson
mass $m_h \ge 114.3$~GeV.

The small values of  $m_{1/2}$ do not satisfy also the constraint coming from the rate of
the rare decay $BR(b\to X_s\gamma)$. However, in this case the dependence on $\tan\beta$
is opposite to that of the previous case. In the case of $\tan\beta=35$ only a small part
of the parameter space is forbidden ($m_0,m_{1/2}\preceq300$~GeV), while for the large
value of $\tan\beta=50$ the forbidden region is  $m_{1/2} \preceq 300 \div 400$~GeV for
any values of $m_0$. This constraint happens to be more restrictive for large $\tan\beta$
than the one related to the Higgs boson mass.

Constraint on experimental value of the anomalous magnetic moment of muon leaves the
allowed band in the   $(m_0,m_{1/2})$ plane the width of which depends on $\tan\beta$.
For $\tan\beta=35$  the excluded regions are the left low corner  ($m_0,m_{1/2}\preceq
300$~GeV (as for the $BR(b\to s\gamma)$ constraint) and almost all right upper part of
the plane. In this case we get a restriction from above on the masses of supoerpartners.
For $\tan\beta=50$ small values of $m_0$ and $m_{1/2}$ are also excluded, however, the
upper bound is shifted to the region above 1~TeV, allowing heavier superpartners.

In the case of $R$-parity conservation the lightest superparticle (LSP) is usually the
neutralino, a certain mixture of superpartners of the photon, $Z$-boson and neutral Higgs
bosons. And it is stable! However, in large amount of the parameter space (left upper
corner of the plane where $m_0<m_{1/2}$) the superpartner of $\tau$-lepton becomes even
lighter and hence have to be stable. But we would have registered a stable charged
particle if it exists even if it is heavy.  Therefore the requirement of neutrality of
LSP have to be fulfilled. The region which is excluded by this requirement also depends
on $\tan\beta$: the larger is $\tan\beta$ the larger is the excluded region.

The allowed regions of the parameter space in the plane $(m_0,m_{1/2})$ which are left
after taking into account all mentioned constraints are shown in Fig.~\ref{fig:region}
for two values of  $\tan\beta=35$ and $\tan\beta=50$.

Constraint on the amount of the Dark matter in the interval $\Omega h^2 = 0.1 \div 0.3$,
in its turn, cuts out narrow bands in the $(m_0,m_{1/2})$ plane as is shown in
Fig.~\ref{fig:astro} for two values of $\tan\beta$.

With account of recent precise data from  WMAP collaboration (Wilkinson Microwave
Anisotropy Probe)~\cite{WMAP} one has much more severe constraints.
As a result the allowed regions in the  $(m_0,m_{1/2})$ plane are along the narrow band
shown in Fig.~\ref{fig:WMAP}~\cite{Wband}. Note that to satisfy WMAP requirement one
prefers large values of  $\tan\beta \approx 50$. Another comment concerns the importance
of the upper limit coming from WMAP data, the lower limit may be influenced by the other
unknown particles and invisible macro objects.

The constrained MSSM possesses already high predictive power. In the regions of parameter
space where there is no contradiction with experimental data or theoretical requirements
one can get the mass spectra of superpartners and the Higgs bosons and to indicate the
possibilities if their experimental search.

\section{Possible scenaria}

The relation between  $m_0$ and $m_{1/2}$ inside  varies along the
the narrow allowed band. Respectively vary the mass spectrum of
superpartners, the dominance of different creation and decay
processes, the values of the cross sections, the methods of analysis
of experimental data at LHC. Remind that part of parameters are
practically fixed in a sense that one can choose their values to
satisfy the imposed constraints with maximal probability.

Consider several cosmologically acceptable and phenomenologically different regions along
the WMAP band.  They have a certain mass spectrum typical to the each region that defines
the main production and annihilation and/or co-annihilation channels for the
neutralino~\cite{regions}.
\begin{itemize}
\item
The first, mostly studied region is the \textit{bulk annihilation region}, this is the
region of relatively small $m_0$ and $m_{1/2}$ ($m_0 \approx 50 \div 150$~GeV,
$m_{1/2}\approx 50 \div 350$~GeV). It is bounded from below by the non-observation of the
Higgs boson and the absence of radiative electroweak symmetry breaking as well as by
consistency with the $b\to s\gamma$ decay rate. From the left there is a forbidden region
where stau is the LSP.

One of the main processes in this region is the annihilation of pair of neutralinos into
quarks through the exchange of a squark in the $t$-channel $\tilde\chi^0_1 \tilde\chi^0_1
\to q \bar{q}$. The parameters can be adjusted in a way to give the right amount of the
Dark matter.

The size of the region depends on $\tan\beta$ and for low values of $\tan\beta$ it
practically disappears due to the non-observation of the Higgs boson.
\item
The other interesting region is the so-called \textit{stau co-annihilation} region. Here
typically one has small values of  $m_0$ and much bigger values of  $m_{1/2}$. It is
located along the border line between the regions where $\tilde\tau_1$-slepton is the LSP
and neutralino $\tilde\chi^0_1$ is the LSP. Evidently this corresponds to the case when
the particles are almost degenerate in masses $m_{\tilde\chi^0_1} \approx
m_{\tilde\tau_1}$ and in the early Universe there were co-annihilation processes
$\tilde\chi^0_1 \tilde\tau_1$ ($\tilde\chi^0_1 \tilde\tau_1 \to \tau^* \to \tau\gamma$)
as well as co-annihilation $\tilde\tau_1 \tilde\tau_1$. Neutralino in this case is mostly
higgsino and its mass may be large up to 500~GeV without violating the WMAP bound.

Co-annihilation region is interesting from the point of view of existence of long-lived
charged sleptons. Their life-time may be large enough to be produced in proton-proton
collisions and to fly away from the detector area or to decay inside the detector at a
considerable distance from the collision point. Clearly that such an event can not be
unnoticed. However, to realize this possibility one need a fine-tuning of the parameters
of the model~\cite{manuel}.
\item
As has been already mentioned for large $m_0$ small values of
$m_{1/2}$ are forbidden due to the absence of radiative electroweak
symmetry breaking. However, along the border of the forbidden
region, the WMAP allowed band may stay long enough leading to the
masses of squarks and sleptons up to a few TeV. This region is
called the \textit{focus point region} since the values of the Higgs
mass parameters here tend to the focus point when running the
renormalization group equations. In this region the Higgs mixing
parameter $\mu$ happens to be small $|\mu|\sim M_Z$. Then it is
possible that two light neutralinos and the light chargino are
practically degenerate $m_{\chi^0_1} \sim m_{\chi^0_2} \sim
m_{\chi^{\pm}_1} \sim \mu$. The lightest neutralino in this case is
mostly higgsino.  The main annihilation channel is the one into the
pair of gauge bosons $\tilde\chi^0_1 \tilde\chi^0_1 \to ZZ$ or
$\tilde\chi^0_1 \tilde\chi^0_1 \to W^+W^-$ but due to degeneracy of
masses of neutralino and chargino there are also possible the
co-annihilation processes $\chi^0_1 \chi^{\pm}_1$, $\chi^0_1
\chi^0_2$, $\chi^+_1 \chi^-_1$ and $\chi^0_2 \chi^{\pm}_1$.

Despite the large values of  $m_{1/2}$ up to  1~TeV, $\mu$ remains
small and leads to chargino and neutralino massses of the order of a
few hundreds GeV. This tells us that the focus point region is
accessible by LHC. Even the cross section of gluino pair production
is big enough to observe this process.
\item
For large values of $\tan\beta$ it is possible that  $m_A \approx 2
m_{\tilde\chi^0_1}$. There is no need for precise equality because
the width of the $CP$-odd Higgs boson $A$ is about tens of GeV. In
this region of  $(m_0,m_{1/2})$ plane the allowed WMAP band has a
sharp turn in the form of a funnel: $A$-\textit{funnel region}. The
main channel of annihilation in this region is  $\tilde\chi^0_1
\tilde\chi^0_1 \to A \to b \bar{b}$ or $\tau \bar\tau$. The reason
for such a behaviour is that for increasing  $\tan\beta$ the mass of
a pseudoscalar Higgs boson  $A$ decreases while the mass of
neutralino practically does not change. Then inevitably  the
resonance situation when  $m_A = 2 m_{\tilde\chi^0_1}$ occurs. And
despite that fact that neutralino in this case is almost photino and
does not interact with the Higgs boson $A$, the tiny admixture of
higgsino leads to considerable effect due to relatively big coupling
of the $A$-boson to quarks and leptons $A b \bar{b}$ è $A \tau
\bar\tau$. For the same reason the exchange of the heavy Higgs boson
$H$ might give an essential contribution.

Besides, in this region the cross section of neutralino $\chi^0_1$ scattering on the
nucleus is of the order of  $10^{-8} \div 10^{-9}$~pb which is close to the values
corresponding to the sensitivity of the modern and the nearest future experiments on the
direct Dark matter searches.
\end{itemize}

In addition to the above mentioned regions there are some small
exotic ones. For example, for a specific choice of parameters (very
big  $A_0$, moderate or big $m_0$ and small $m_{1/2}$) as a result
of mixing one of the $t$-squarks becomes practically degenerate with
the lightest neutralino $\chi^0_1$. In this case the process of
$\tilde\chi^0_1 \tilde t_1$ co-annihilation is possible. For small
values of $m_{1/2}$ (and for appropriate choice of the other
parameters) there is a possibility of neutralino annihilation due to
light Higgs boson exchange in the $s$-channel. This situation is
analogous to that of annihilation through $A$ or $H$.

\begin{itemize}
\item
One should mention the other interesting constraint on the parameter
space of the MSSM related to the supersymmetric interpretation of
the excess of the diffuse gamma ray flux in our Galaxy compared to
the background calculations. This is the data presented by EGRET
collaboration (Energetic Gamma Ray Experiment
Telescope)~\cite{egret1}. Omitting the details we notice, that it is
enough to assume the existence of a neutral stable weakly
interacting particle (WIMP) of a certain mass to explain this
excess.  The fit gives the value of this mass in the region $m_X
\approx 50-100$~GeV~\cite{egret2}. If one takes this WIMP to be the
lightest neutralino this will strongly constrain the value of
parameter $m_{1/2}$. Moreover, this constraint is compatible with
WMAP. The allowed area is in the  region of $m_0 \approx 1400$~GeV
and $m_{1/2}\approx 180$~GeV~\cite{egret3}, i.e. practically between
the bulk annihilation region and then focus point region.
\end{itemize}

\section{Search for supersymmetry at LHC}

The starategy of SUSY searches at LHC is based on the assumption
that the masses of superpartners indeed are in the region of  1~TeV
so that they might be created at the mass shell with the cross
section big enough to distinguish them from the background of the
ordinary particles. Calculation of the background in the framework
of the Standard Model thus becomes essential since the secondary
particles in all the cases will be the same.

There are many possibilities to create superpartners at hadron
colliders. Besides the usual annihilation channel there are numerous
processes of gluon fusion, quark-antiquark and quark-gluon
scattering. The maximal cross sections of the order of a few
picobarn can be achieved in the process of gluon fusion.

As a rule all superpartners are short lived and decay into the ordinary particles and the
lightest superparticle.  The main decay modes of superpartners, i.e. experimental
manifestation of SUSY at LHC are presented  in Table~\ref{tab:modes}.
\begin{table}[htb]\begin{center}
\caption{Creation of superpartners and the main decay modes} \label{tab:modes}
\bigskip
\begin{tabular}{|l|l|l|} \hline
~~Creation & The main decay modes & ~~~~~~~Signature\\
\hline
~$\bullet
~~\tilde{g}\tilde{g}, \tilde{q}\tilde{q}, \tilde{g}\tilde{q}$ &
$\left.\begin{array}{l} \tilde{g} \to q\bar q \tilde{\chi}^0_1   \\
 ~~~~~ q\bar q' \tilde{\chi}^\pm_1  \\
 ~~~~~ g\tilde{\chi}^0_1 \end{array} \right\}
 m_{\tilde{q}}>m_{\tilde{g}}$ &
$\begin{array}{c} \Big/ \hspace{-0.3cm E_T}
+ \mbox{~multijets}~(+\mbox{leptons}) \end{array}$ \\
& $\left.\begin{array}{l}\tilde{q} \to q \tilde{\chi}^0_i \\
    \tilde{q} \to q' \tilde{\chi}^\pm_i \end{array} \right\}
  m_{\tilde{g}}>m_{\tilde{q}}$ & \\ \hline
~$\bullet
~~\tilde{\chi}^\pm_1\tilde{\chi}^0_2$ & ~$\tilde{\chi}^\pm_1
\to \tilde{\chi}^0_1 \ell^\pm \nu, \ \tilde{\chi}^0_2 \to
 \tilde{\chi}^0_1 \ell\ell$ &
$\mbox{~trilepton} + \Big/ \hspace{-0.3cm E_T}$ \\
 & ~$\tilde{\chi}^\pm_1 \to \tilde{\chi}^0_1 q \bar q',
\tilde{\chi}^0_2 \to \tilde{\chi}^0_1 \ell\ell,$ &
$\mbox{~dileptons + jet} + \Big/ \hspace{-0.3cm E_T}$ \\
\hline
~$\bullet
~~\tilde{\chi}^+_1\tilde{\chi}^-_1$ &
~$\tilde{\chi}^+_1 \to \ell \tilde{\chi}^0_1 \ell^\pm \nu$ &
$\mbox{~dilepton} + \Big/ \hspace{-0.3cm E_T}$ \\
\hline
~$\bullet
~~\tilde{\chi}^0_i\tilde{\chi}^0_i$ &
~$\tilde{\chi}^0_i \to \tilde{\chi}^0_1 X, \tilde{\chi}^0_i \to \tilde{\chi}^0_1 X'$ &
$\mbox{~dilepton+jet} + \Big/ \hspace{-0.3cm E_T}$ \\
\hline
~$\bullet
~~\tilde{t}_1\tilde{t}_1$ &
$~\tilde{t}_1 \to c \tilde{\chi}^0_1$ &
$\mbox{~2 noncollinear jets}+ \Big/\hspace{-0.3cm E_T}$ \\
 & $~\tilde{t}_1 \to b \tilde{\chi}^\pm_1, \tilde{\chi}^\pm_1 \to \tilde{\chi}^0_1 q\bar q'$ &
$\mbox{~single lepton} + \Big/ \hspace{-0.3cm E_T} + b's$ \\
 &~$\tilde{t}_1 \to b \tilde{\chi}^\pm_1,\tilde{\chi}^\pm_1 \to \tilde{\chi}^0_1 \ell^\pm\nu,$ &
$\mbox{~dilepton} + \Big/ \hspace{-0.3cm E_T} + b's $ \\
\hline
~$\bullet
~~\tilde{l}\tilde{l},\tilde{l}\tilde{\nu},\tilde{\nu}\tilde{\nu}$ &
 $~\tilde{\ell}^\pm \to \ell^\pm \tilde{\chi}^0_i,\tilde{\ell}^\pm \to \nu_\ell \tilde{\chi}^\pm_i$ &
 $\mbox{~dilepton}+ \Big/ \hspace{-0.3cm E_T}$ \\
& $~\tilde{\nu} \to \nu \tilde{\chi}^0_1$ & $\mbox{~single~lepton} + \Big/ \hspace{-0.3cm
E_T}$
\\ \hline
\end{tabular}\end{center}
\end{table}

Notice the typical events with missing energy and transverse momentum that is the main
difference from the background processes of the Standard Model. The missing energy is
carried away by the heavy particle with the mass of the order of 100~GeV that is
essentially different from the processes with neutrino in the final state.
\begin{table}[htb]
\caption{Creation of the pair of gluino with further cascade decay} \label{tab:n}
\bigskip
\begin{tabular}{|c|p{0.9cm}|}
\hline Process & final \\
& states \\ \hline
\includegraphics[width=50mm,height=40mm]{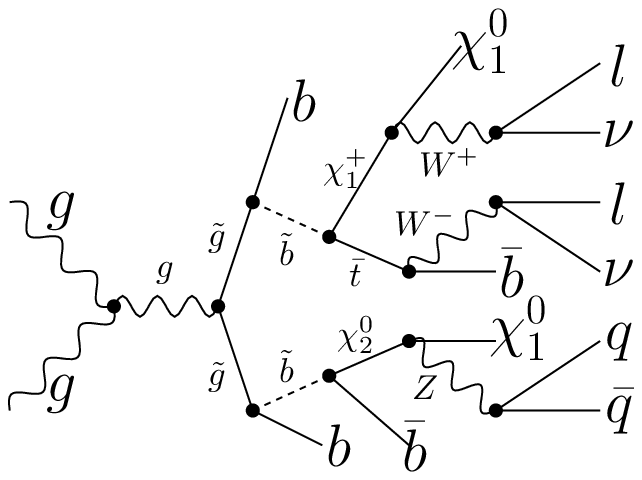}
& \vspace*{-35mm}
\begin{minipage}[t]{1.8cm}
$\begin{array}{c}
2\ell \\ 2\nu \\ 6j \\ \Big/\hspace{-0.3cm E_T}
\end{array}$
\end{minipage} \\ \hline
\includegraphics[width=50mm]{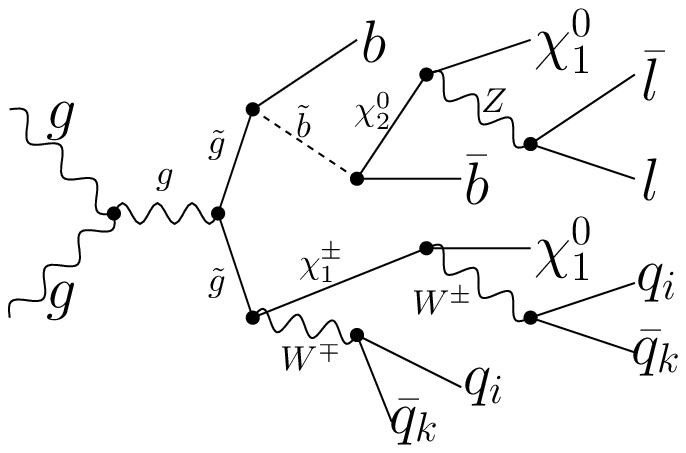}
& \vspace*{-30mm}
\begin{minipage}[t]{1.8cm}
$\begin{array}{c}
2\ell \\ 6j \\ \Big/ \hspace{-0.3cm E_T}
\end{array}$
\end{minipage} \\ \hline
\includegraphics[width=50mm]{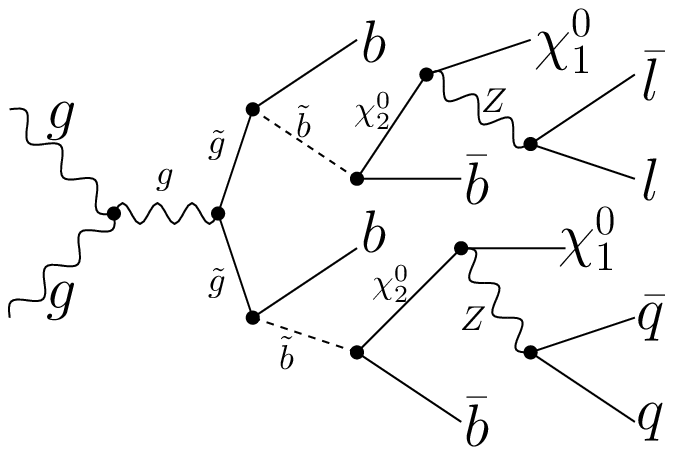}
& \vspace*{-30mm}
\begin{minipage}[t]{1.8cm}
$\begin{array}{c}
2\ell \\ 6j \\ \Big/ \hspace{-0.3cm E_T}
\end{array}$
\end{minipage} \\ \hline
\end{tabular}
\hspace*{0.9mm}
\begin{tabular}{|c|p{0.9cm}|}
\hline Process & final \\
& states \\ \hline
\includegraphics[width=50mm,height=40mm]{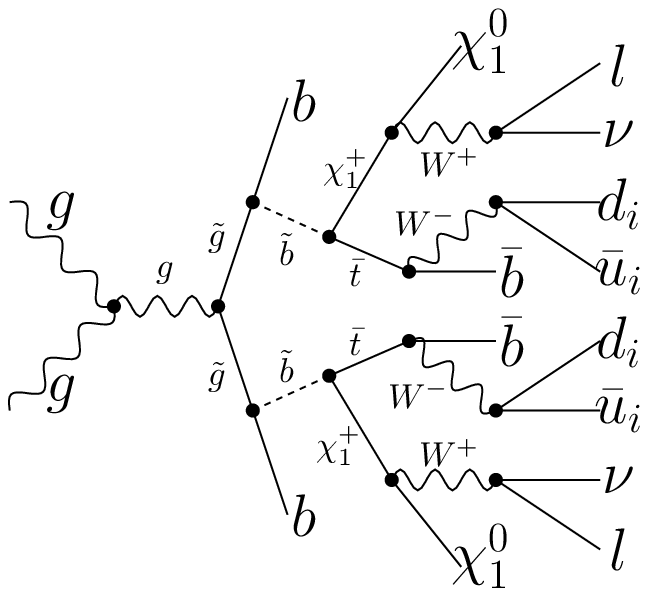}
& \vspace*{-35mm}
\begin{minipage}[t]{1.8cm}
$\begin{array}{c}
2\ell \\ 2\nu \\ 8j \\ \Big/\hspace{-0.3cm E_T}
\end{array}$
\end{minipage} \\ \hline
\includegraphics[width=50mm]{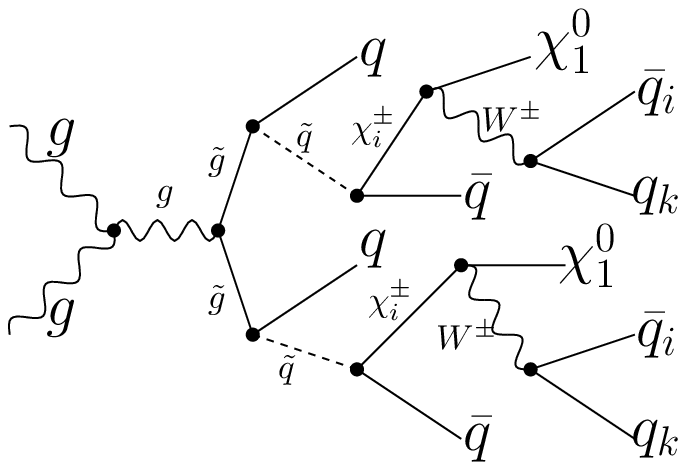}
& \vspace*{-25mm}
\begin{minipage}[t]{1.8cm}
$\begin{array}{c}
8j \\ \Big/ \hspace{-0.3cm E_T}
\end{array}$
\end{minipage} \\ \hline
\includegraphics[width=50mm]{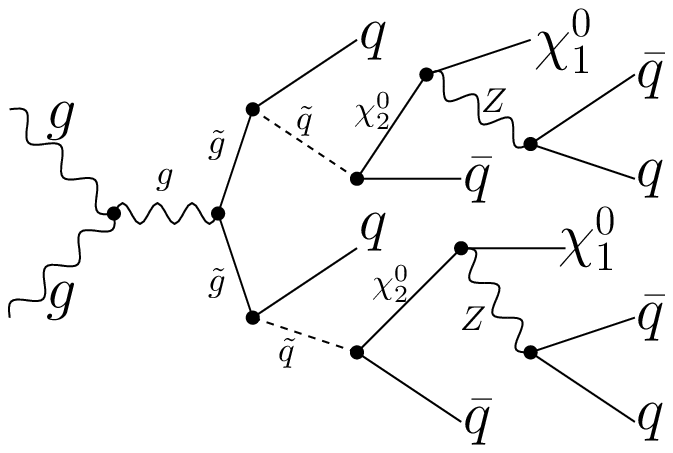}
& \vspace*{-25mm}
\begin{minipage}[t]{1.8cm}
$\begin{array}{c}
8j \\ \Big/ \hspace{-0.3cm E_T}
\end{array}$
\end{minipage} \\ \hline
\end{tabular}
\end{table}
In hadron collisions the superpartners are always created in pairs and then further
quickly decay creating a cascade with the ordinary quarks (i.e. hadron jets) or leptons
at the end plus the missing energy. For the case of gluon fusion with creation of gluino
it is presented in Table~\ref{tab:n}.

Chargino and neutralino can also be produced in pairs through the
Drell-Yang mechanism $pp \to \tilde \chi^\pm_1 \tilde \chi^0_2$ and
can be detected via their lepton decays $\tilde \chi^\pm_1 \tilde
\chi^0_2 \to \ell\ell\ell+\Big/ \hspace{-0.3cm E_T}$. Hence the main
signal of their creation is the isolated leptons and missing energy
(Table~\ref{tab:j}). The main background in trilepton channel comes
from creation of the standard particles  $WZ/ZZ,t\bar t, Zb\bar b$
and $b\bar b$. There might be also the supersymmetric background
from the cascade decays of squarks and gluino into multilepton
modes.

\begin{table}[htb]
\caption{Creation of the lightest chargino and the second neutralino with further cascade
decay.} \label{tab:j}
\bigskip
\begin{tabular}{|c|p{0.9cm}|} \hline
Process & final \\ & states \\
\hline
\includegraphics[width=50mm,height=40mm]{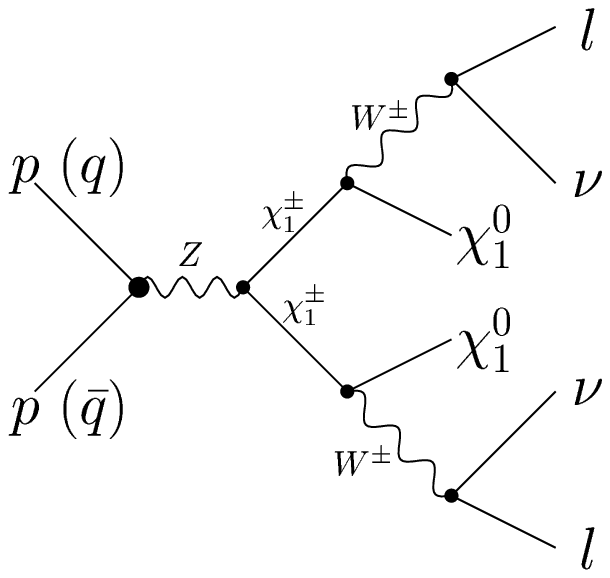}
& \vspace*{-30mm}
\begin{minipage}[t]{1.8cm}
$\begin{array}{c}
2\ell \\ 2\nu \\ \Big/\hspace{-0.3cm E_T}
\end{array}$
\end{minipage} \\ \hline
\includegraphics[width=50mm,height=40mm]{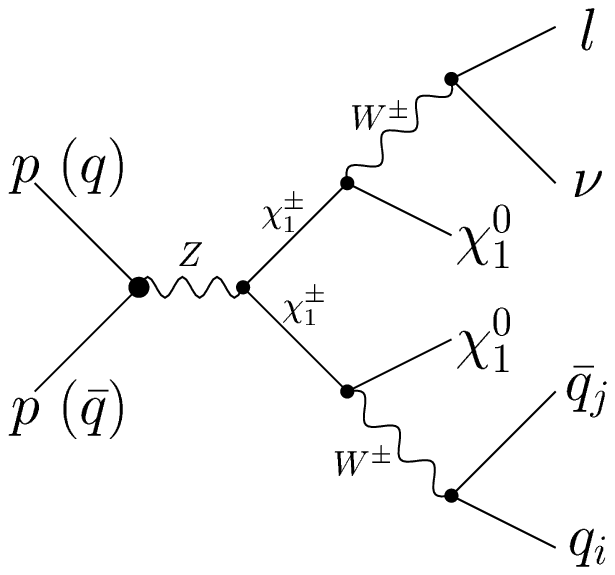}
& \vspace*{-35mm}
\begin{minipage}[t]{1.8cm}
$\begin{array}{c}
\ell \\ \nu \\ 2j \\ \Big/\hspace{-0.3cm E_T}
\end{array}$
\end{minipage} \\ \hline
\includegraphics[width=50mm,height=40mm]{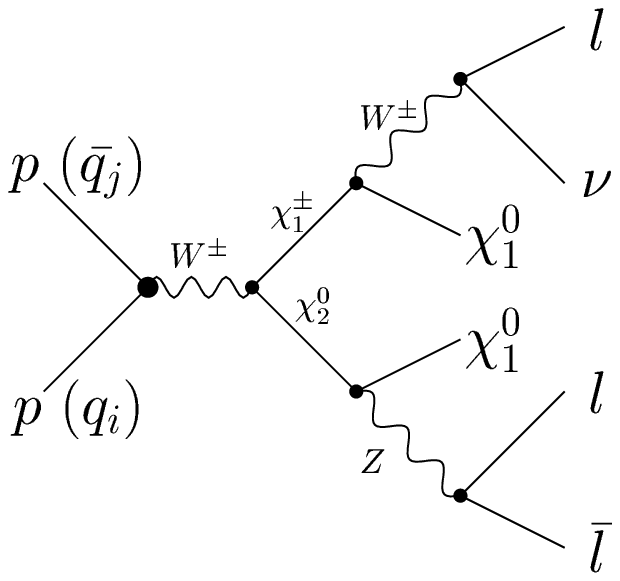}
& \vspace*{-30mm}
\begin{minipage}[t]{1.8cm}
$\begin{array}{c}
3\ell \\ \nu \\ \Big/\hspace{-0.3cm E_T}
\end{array}$
\end{minipage} \\ \hline
\end{tabular}
\hspace*{0.9mm}
\begin{tabular}{|c|p{0.9cm}|} \hline
Process & final\\ & states \\
\hline
\includegraphics[width=50mm,height=40mm]{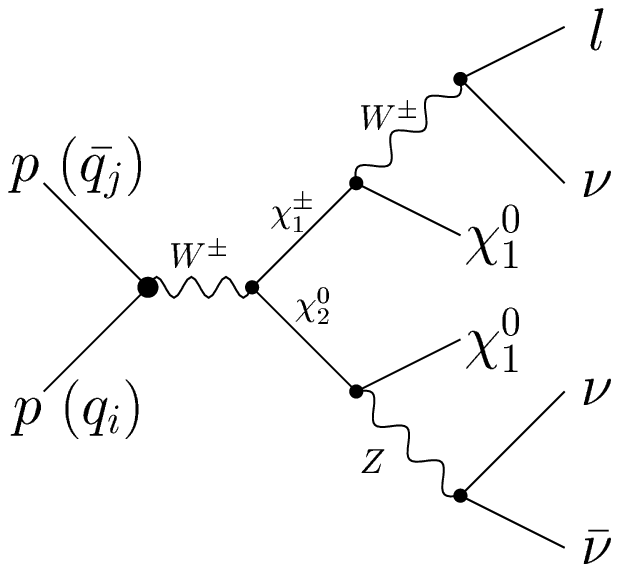}
& \vspace*{-30mm}
\begin{minipage}[t]{1.8cm}
$\begin{array}{c}
\ell \\ 3\nu \\ \Big/\hspace{-0.3cm E_T}
\end{array}$
\end{minipage} \\ \hline
\includegraphics[width=50mm,height=40mm]{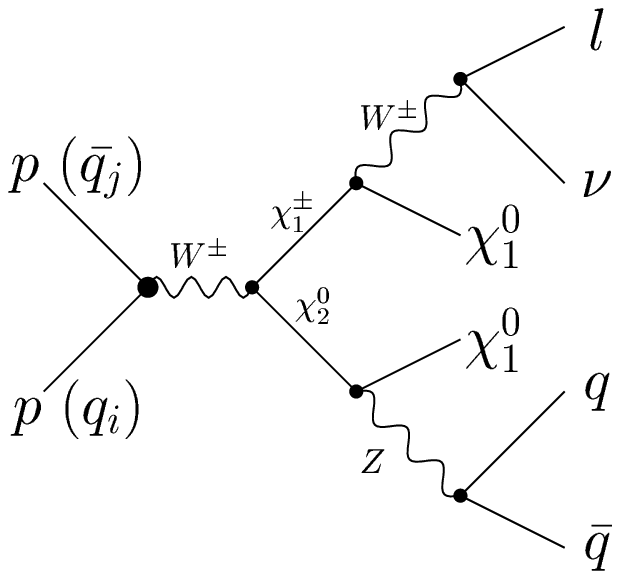}
& \vspace*{-35mm}
\begin{minipage}[t]{1.8cm}
$\begin{array}{c}
\ell \\ \nu \\ 2j \\ \Big/\hspace{-0.3cm E_T}
\end{array}$
\end{minipage} \\ \hline
\includegraphics[width=50mm,height=40mm]{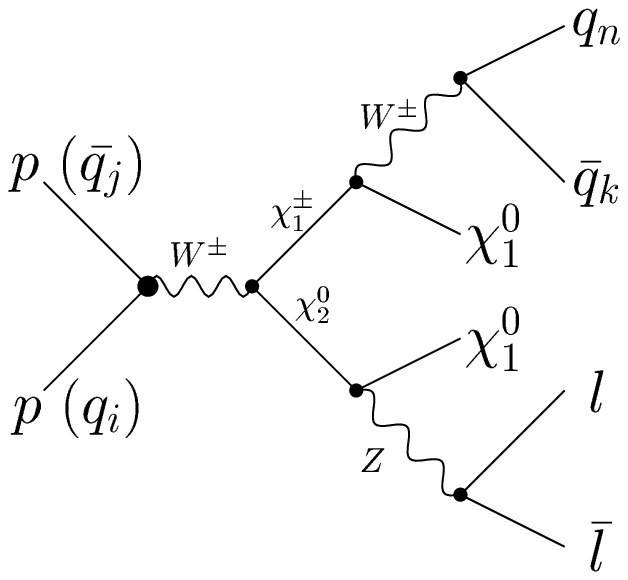}
& \vspace*{-30mm}
\begin{minipage}[t]{1.8cm}
$\begin{array}{c}
2\ell \\ 2j \\ \Big/\hspace{-0.3cm E_T}
\end{array}$
\end{minipage} \\ \hline
\end{tabular}
\end{table}

\begin{figure}[htb]
\begin{center}
\includegraphics[width=0.40\textwidth]{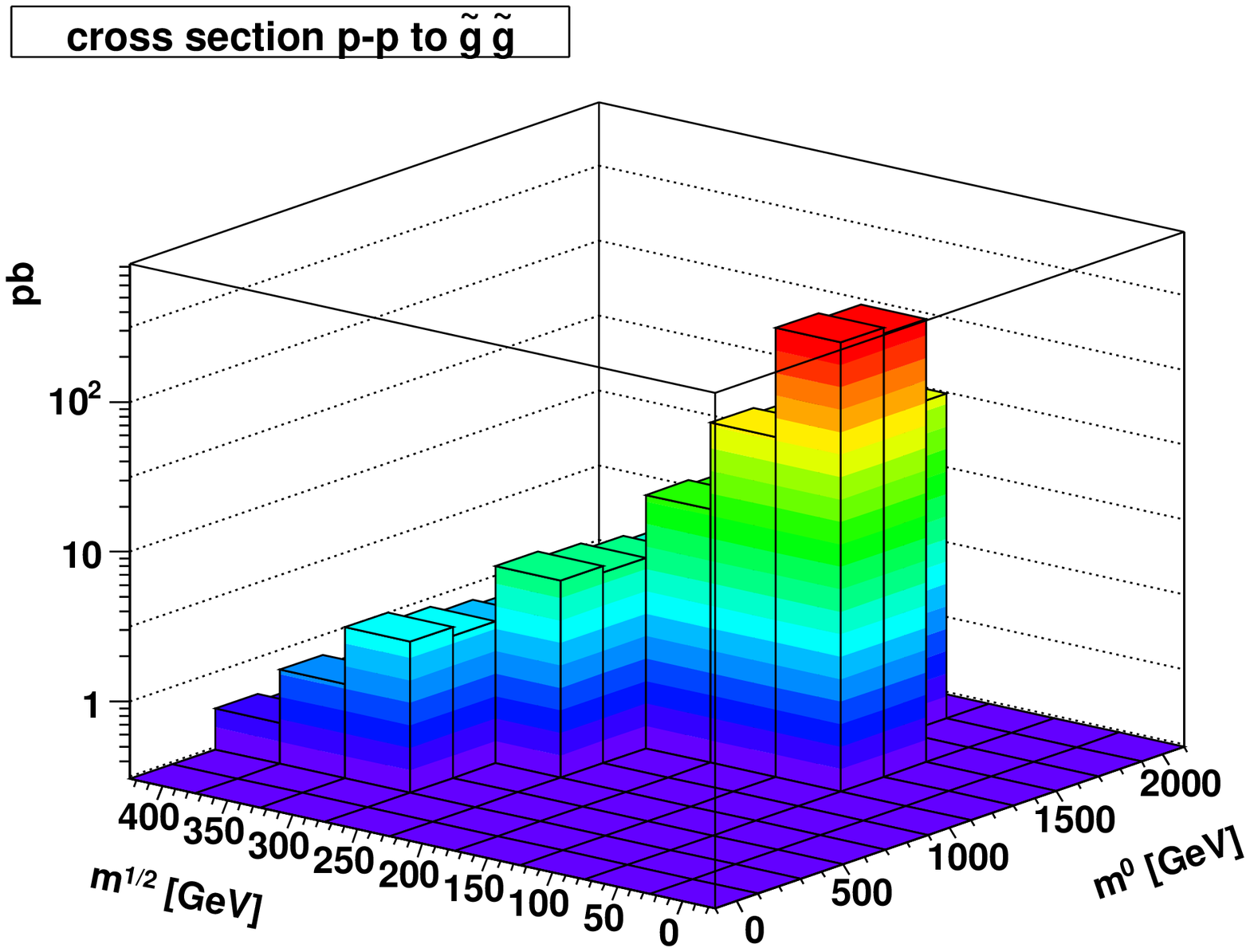}
\includegraphics[width=0.40\textwidth]{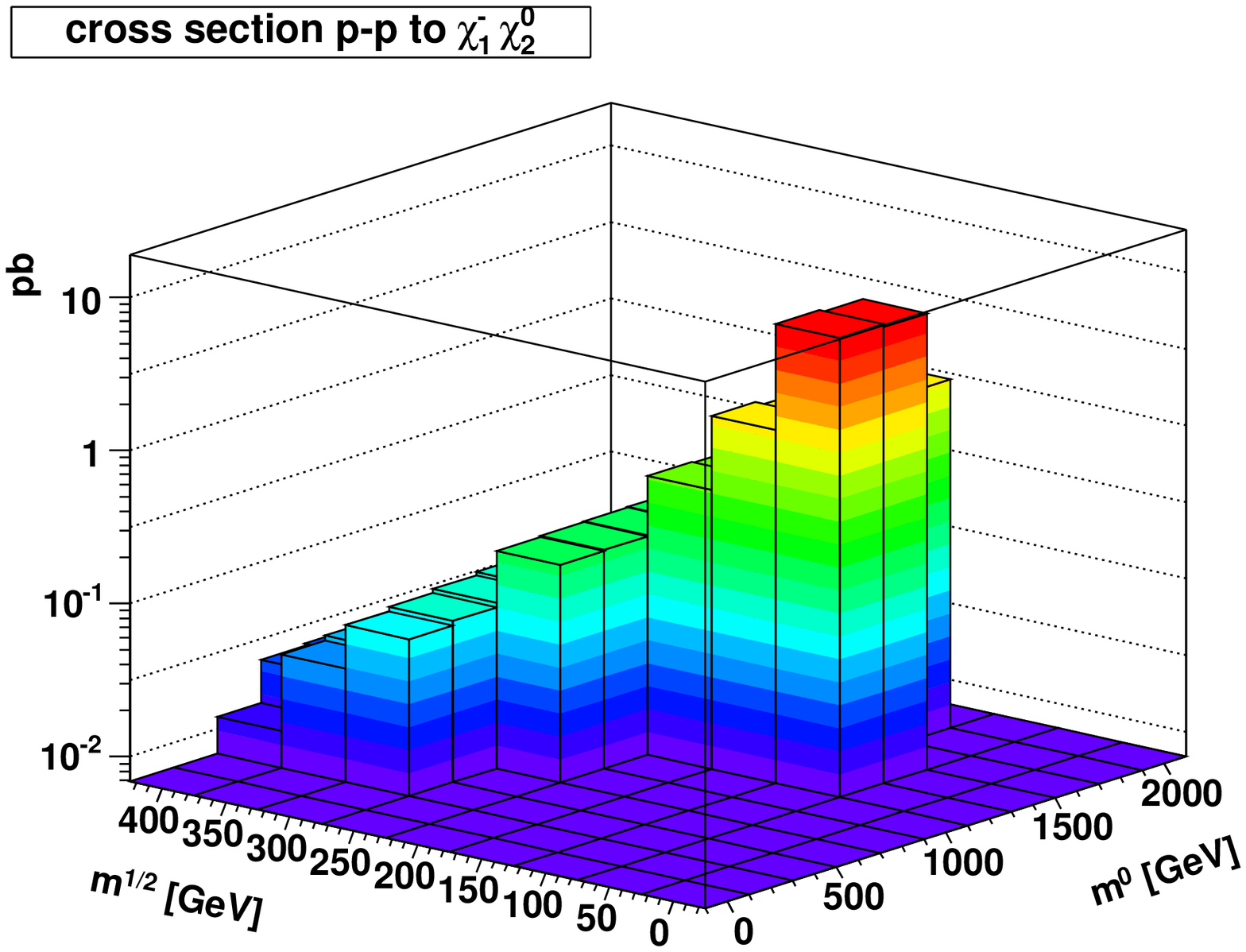}
\includegraphics[width=0.40\textwidth]{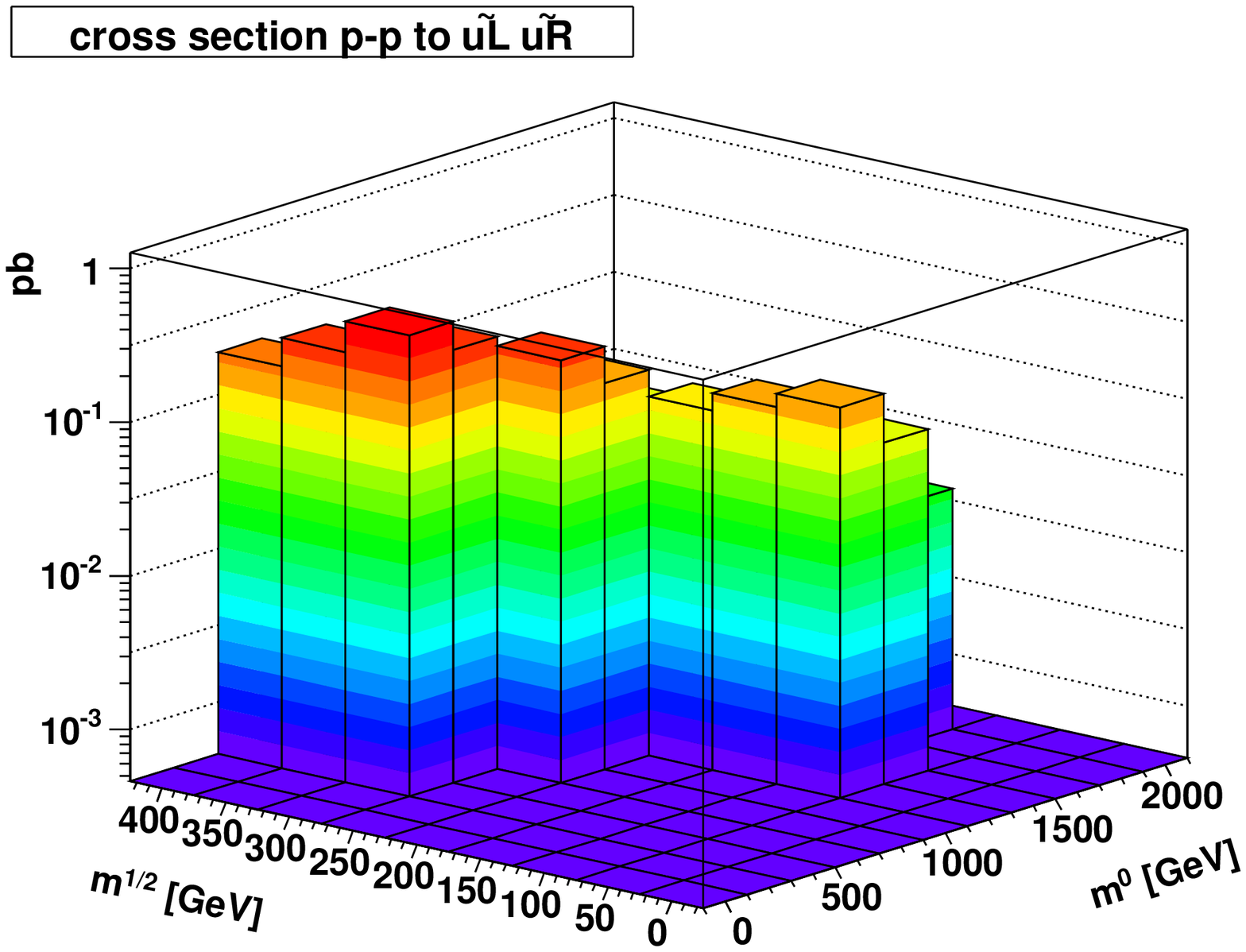}
\includegraphics[width=0.40\textwidth]{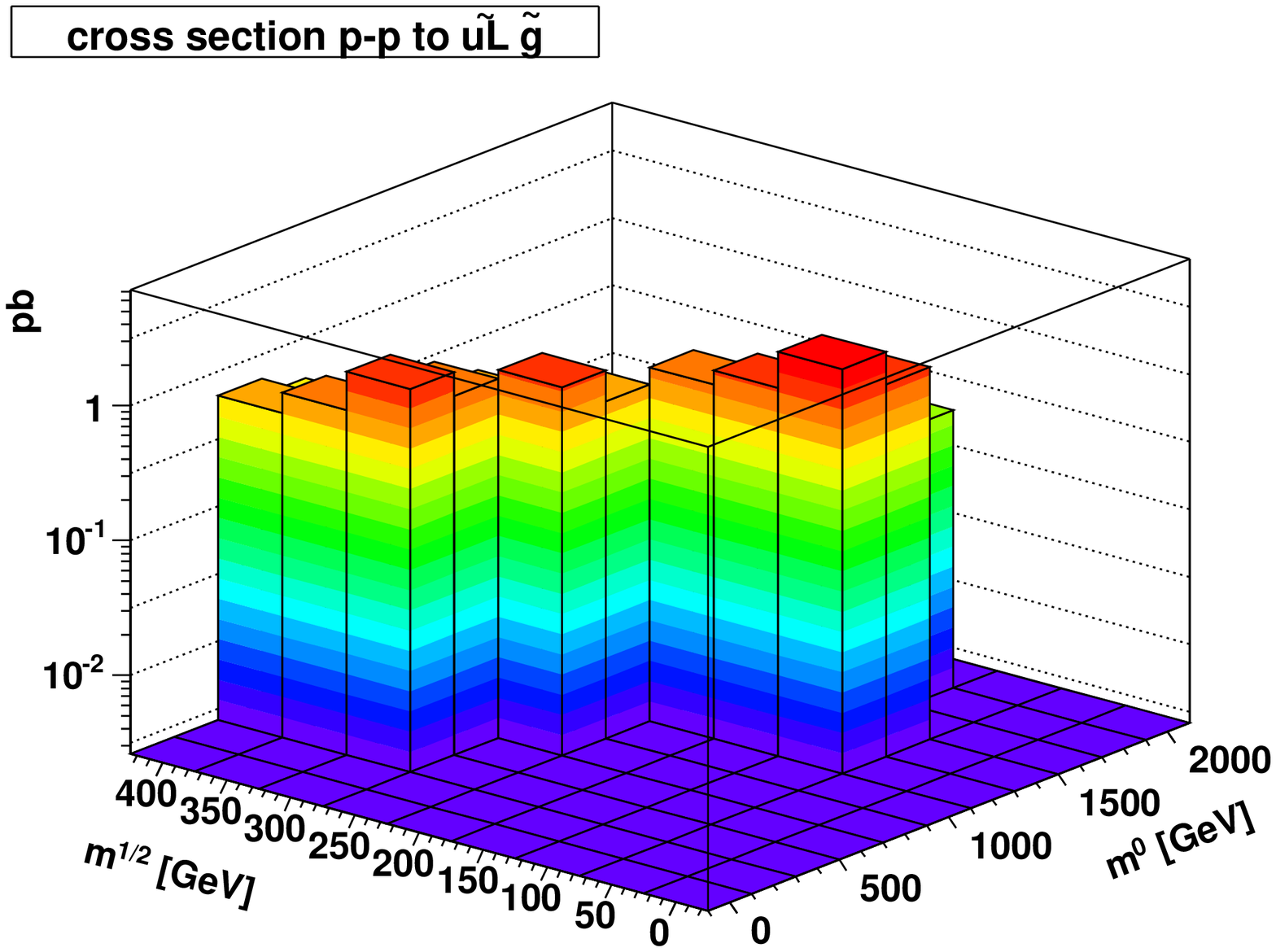}
\end{center}
\caption{The cross sections of superpartners creation as functions of $m_{1/2}$ and
$m_{0}$ for $\tan\beta=51$, $A_0=0$ and positive sign of $\mu$.} \label{fig:Mss}
\end{figure}

The cross sections for various superpartners creation at LHC are shown in
Fig.~\ref{fig:Mss}. One can see that in some regions they may reach a few pb that is for
a planned luminosity of LHC allows one to provide reliable detection. In the case of
light neutralino and chargino  the cross sections of their pair production can reach
those of the strongly interacting particles~\cite{unexpLHC}.
%
\begin{figure}[hb]
\begin{center}
\includegraphics[width=0.35\textwidth]{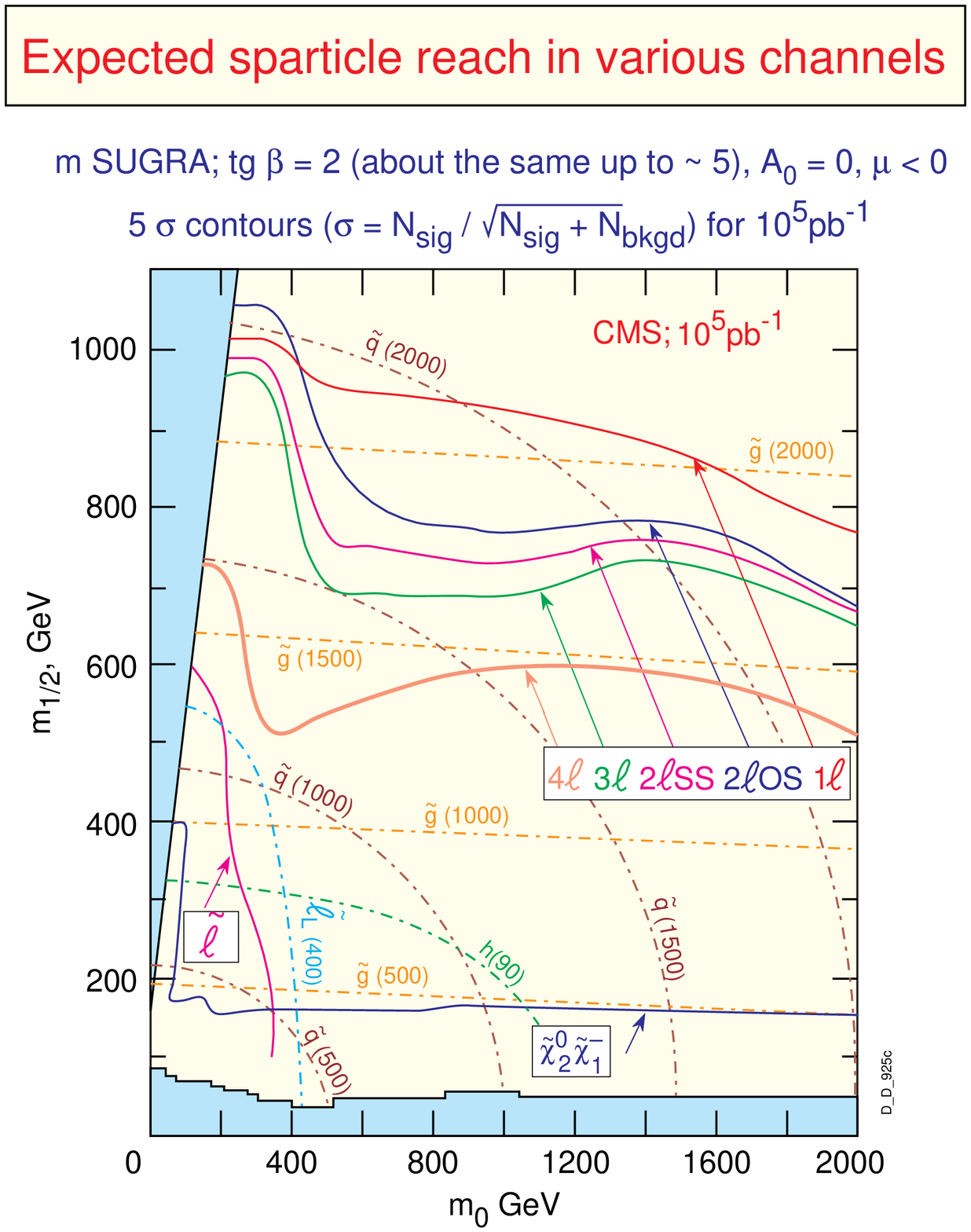}
\includegraphics[width=0.35\textwidth]{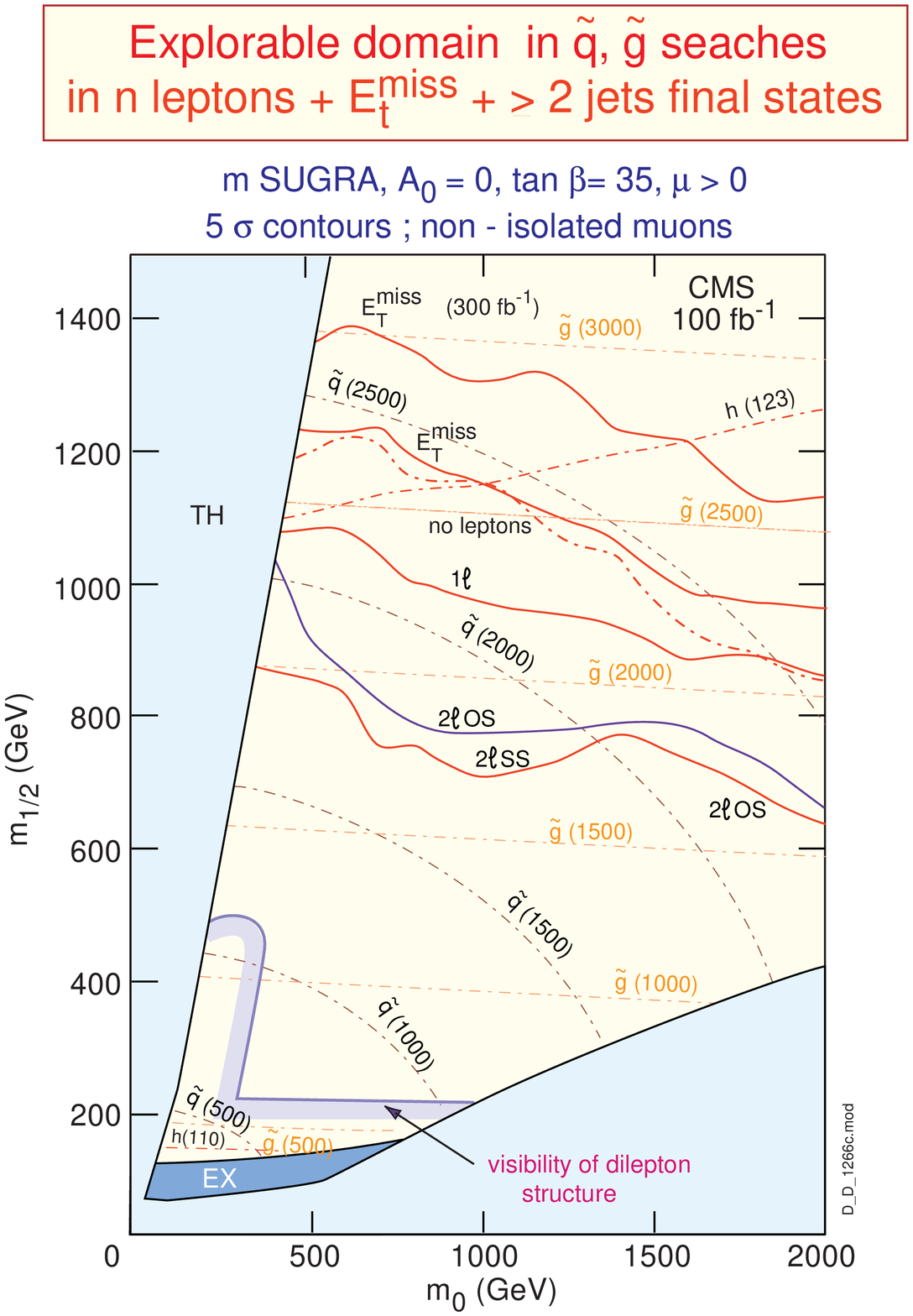}
\end{center}
\caption{Expected renge of reach for superpartners in various channels at
LHC~\cite{LHC}.} \label{fig:LHC1}
\end{figure}
To present the region of reach for the LHC in different channels of
sparticle production one uses the same plane of soft SUSY breaking
parameters $m_0$ and $m_{1/2}$. In this case one usually assumes
certain luminosity which will be presumably achieved during the
accelerator operation.

Thus, for instance, in Fig.~\ref{fig:LHC1} it is shown the regions of reach in different
channels. The lines of a constant squark mass form the arch curves, and those for gluino
are almost horisontal. The curved lines show the reach bounds in different channel of
creation of secondary particles. The theoretical curves are obtained within the MSSM for
a certain choice of the other soft SUSY breaking parameters. In the left plot the
calculations are performed for the luminosity equal to $10^5$~pb$^{-1}$ and in the right
one for the luminosity $10^2$~pb$^{-1}$ which will be presumably reached at the first
stage. As one can see, for the fortunate circumstances the wide range of the parameter
space  up to the masses of the order of 2~Tev will be examined.

\begin{figure}[b]
\begin{center}
\includegraphics[width=0.35\textwidth]{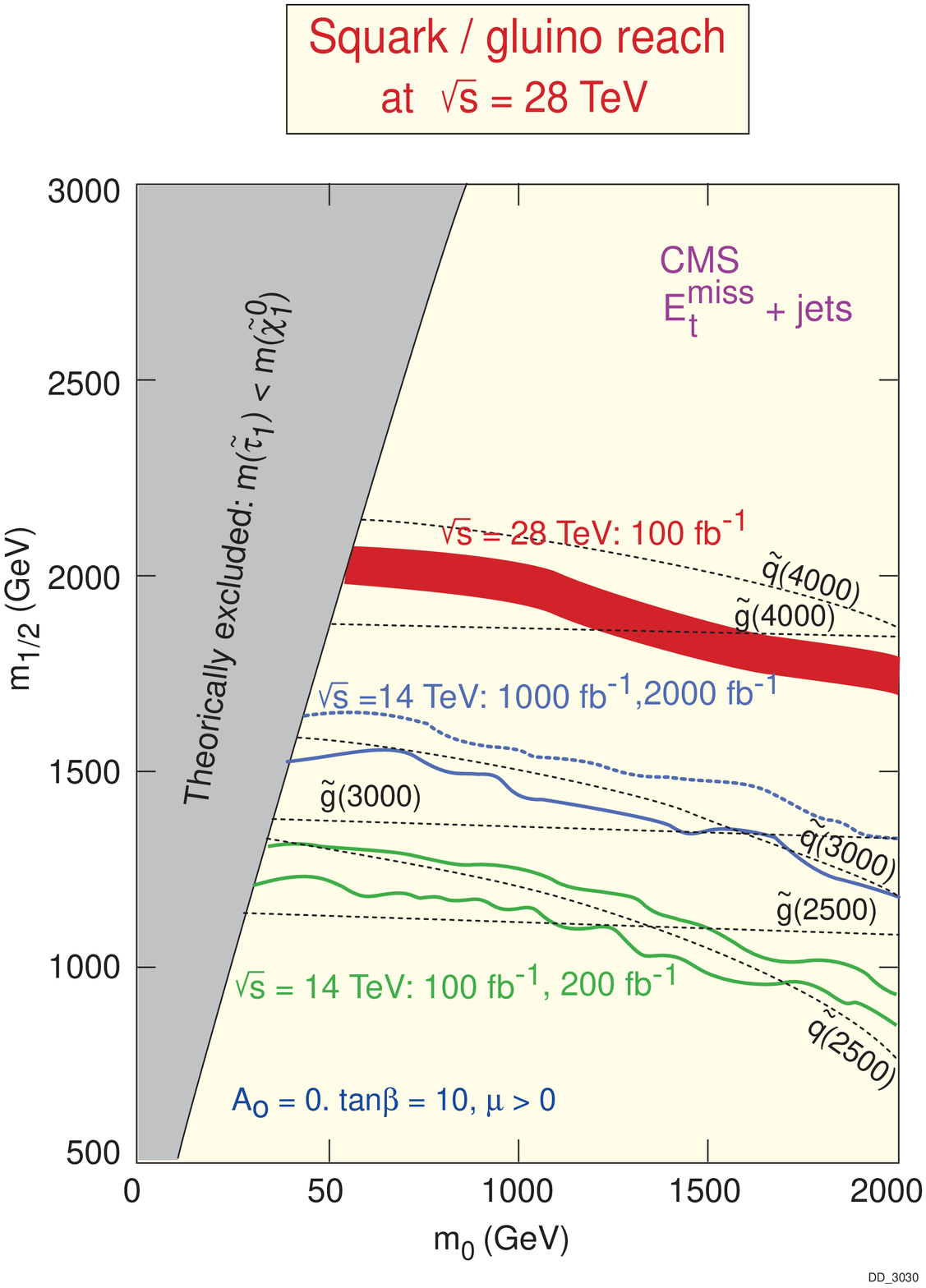}
\includegraphics[width=0.35\textwidth]{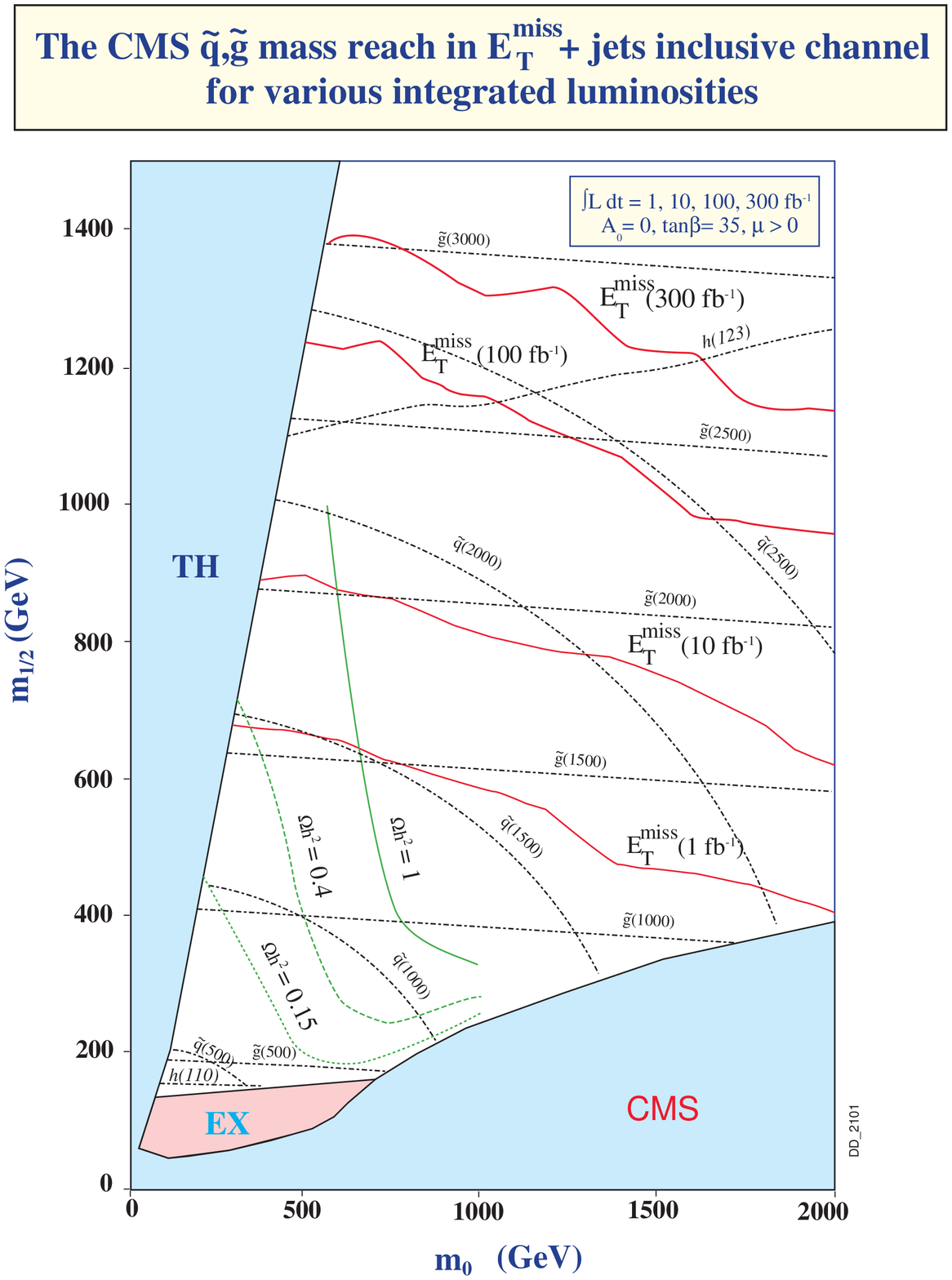}
\end{center}
\caption{Expected range of reach for squarks and gluino for diffewrent luminosities at
LHC~\cite{LHC}.} \label{fig:LHC2}
\end{figure}

The other example is shown in Fig.~\ref{fig:LHC2} where the regions
of reach for squarks and gluino  are shown for various luminosities.
One can see that for the maximal luminosity the discovery range for
squarks and gluino reaches 3~TeV for the center of mass energy of
14~TeV and even higher for the double energy.

\begin{figure}[htb]
\begin{center}
\includegraphics[width=0.7\textwidth]{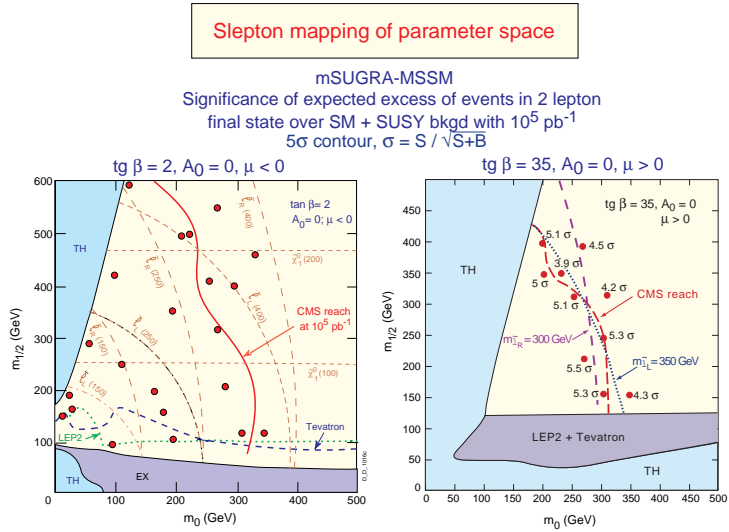}
\end{center}
\caption{Expected range of reach for sleptons at LHC~\cite{LHC}.} \label{fig:LHC3}
\end{figure}

The same is true for the sleptons as shown in Fig.~\ref{fig:LHC3}.
The slepton pairs can be created via the Drell-Yang mechanism
$pp\to\gamma^*/Z^*\to\tilde \ell^+\tilde \ell^-$ and can be detected
through the slepton decays $\tilde \ell \to \ell+\tilde \chi^0_1$.
The typical signal used for slepton detection is the dilepton pair
with the missing energy without hadron jets. For the luminosity of
$L_{tot}=100$~fb$^{-1}$ the LHC will be able to discover sleptons
with the masses up to 400~GeV~\cite{Kras}.

We do not discuss here the different possibilities of detection of
long lived supersymmetric particles, staus or supersymmetric
hadrons. The very existence of these particles requires the fine
tuning of parameters. However, if these particles exist, their decay
inside the detector would give a characteristic signal with creation
of a jet or a charged lepton at a point distinct from the collision
point which might be detected.\
%
%
\clearpage

\section{SUSY Higgs boson}
The presence of an extra Higgs doublet in SUSY model is a novel feature of the theory. In
the MSSM  one has two doublets with the quantum numbers (1,2,-1) and (1,2,1),
respectively:
 \begin{equation}
H_1=\left( \begin{array}{c} H^0_1 \\ H_1^- \end{array}  \right) = \left( \begin{array}{c}
v_1 +\frac{\displaystyle S_1+iP_1}{\sqrt{2}} \\ H^-_1 \end{array}\right), \ \ \
H_2=\left(
\begin{array}{c} H^+_2 \\ H_2^0 \end{array}  \right) = \left(
\begin{array}{c} H^+_2 \\ v_2 +\frac{\displaystyle
S_2+iP_2}{\sqrt{2}}
\end{array} \right),
 \end{equation}
where  $v_i$ are the vacuum expectation values of the neutral components.

Thus, in the MSSM, as actually in any two Higgs doublet model,
 one has  8=4+4=5+3 degrees of freedom. As in the case of the SM, 3 degrees of freedom can be
gauged away, and one is left with five  physical Higgs bosons: two
CP-even neutral, one CP-odd neutral  and two charged ones.

The Higgs potential in the MSSM is totally defined by superpotential ${\cal
W_R}$~(\ref{R}) and the soft terms (\ref{soft2}). Due to the structure of ${\cal W_R}$ it
contributes  only to the mass matrix while the Higgs self-interaction comes from the
interaction with the gauge fields.  The tree level potential is
\begin{eqnarray}
V_{tree}(H_1,H_2)&=&m^2_1|H_1|^2+m^2_2|H_2|^2-m^2_3(H_1H_2+h.c.)
\label{Higpot} \nonumber\\ &+&
\frac{g^2+g^{'2}}{8}(|H_1|^2-|H_2|^2)^2 + \frac{g^2}{2}|H_1^\dagger
H_2|^2,
\end{eqnarray}
where $m_1^2=m^2_{H_1}+\mu^2, m_2^2=m^2_{H_2}+\mu^2$. At the GUT scale
$m_1^2=m^2_2=m_0^2+\mu^2_0, \ m^2_3=-B\mu_0$. Notice that the Higgs self-coupling in
eq.(\ref{Higpot}) is  fixed and  defined by the gauge interactions as opposed to the SM.

 The mass eigenstates are~\cite{susy}:
 $$\begin{array}{l} \left\{
\begin{array}{lllr} G^0 \ &=& -\cos\beta P_1+\sin \beta P_2 , & \ \ \
Goldstone \ boson \ \to Z_0, \\ A \ &=& \sin\beta P_1+\cos \beta P_2 , & \ \ \ \ \ \ \ \
Neutral \ CP=-1 \ Higgs, \end{array}\right.\\  \\ \left\{
\begin{array}{lllr} G^+ &=& -\cos\beta (H^-_1)^*+\sin \beta H^+_2 , &\ \
Goldstone \ boson \ \to W^+, \\ H^+ &=& \sin\beta (H^-_1)^*+\cos \beta H^+_2 , &\ \ \
Charged \ Higgs, \end{array}\right.\\ \\ \left\{
\begin{array}{lllr} h \ &=& -\sin\alpha S_1+\cos\alpha S_2 , & \ \ \ \ \ \
\ SM \ Higgs \ boson \ CP=1, \\ H \ &=& \cos\alpha S_1+\sin\alpha S_2 , & \ \ \ \ \ \ \
Extra \ heavy \ Higgs \ boson , \end{array}\right.
\end{array}$$ where the mixing angle $\alpha $ is
given by $$ \tan 2\alpha = \tan 2\beta \left(\frac{m^2_A+M^2_Z}{m^2_A-M^2_Z}\right).$$

The physical Higgs bosons acquire the following masses \cite{susy}:
 \begin{eqnarray} \mbox{CP-odd
neutral Higgs} \ \ A: && \ \ \ \ \ \ \ \ \ \ \ \ m^2_A = m^2_1+m^2_2, \nonumber \\
\mbox{Charge Higgses} \ \ H^{\pm}: && \ \ \ \ \ \
 \ \ \ \ \ m^2_{H^{\pm}}=m^2_A+M^2_W ,
 \end{eqnarray}
CP-even neutral Higgses \ \ H, h:
\begin{equation} m^2_{H,h}=
\frac{1}{2}\left[m^2_A+M^2_Z \pm \sqrt{(m^2_A+M_Z^2)^2-4m^2_AM_Z^2\cos^22\beta}\right],
\end{equation}
where, as usual,
 $$ M^2_W=\frac{g^2}{2}v^2, \ \
M^2_Z=\frac{g^2+g'^2}{2}v^2 .$$ This leads to the once celebrated SUSY mass relations
\begin{equation}\begin{array}{c} m_{H^{\pm}} \geq M_W, \\[0.2cm]
m_h \leq m_A \leq M_H, \\[0.2cm] m_h \leq M_Z |\cos 2\beta| \leq
M_Z , \\[0.2cm]  m_h^2+m_H^2=m_A^2+M_Z^2.\end{array}\label{bound}
\end{equation}

Thus, the lightest neutral Higgs boson happens to be lighter than
the $Z$-boson, which clearly distinguishes it from the SM one since
not knowing  the mass of the Higgs boson in the SM one has several
indirect constraints leading to the lower boundary of $m_h^{SM} \geq
135 $ GeV~\cite{bound}. However, after including the radiative
corrections, the mass of the lightest Higgs boson in the MSSM,
$m_h$, increases.

These radiative corrections vanish when supersymmetry is not broken and depend on the
values of the soft breaking parameters. The main contribution comes from top (stop)
quarks. Contributions from the other particles are much smaller~\cite{radcorr}. In the
one loop order one has the following modification of  the tree-level relation for the
lightest Higgs mass
\begin{equation}
m_h^2 \approx M_Z^2\cos^2 2\beta + \frac{3g^2 m_t^4}{16\pi^2M_W^2} \log\frac{
\tilde{m}_{t_1}^2\tilde{m}_{t_2}^2}{m_t^4}. \label{rad}
\end{equation}
One finds that the one-loop correction is positive and increases the mass value. Two loop
corrections have the opposite effect but are smaller and result in slightly lower value
of  the Higgs mass ~\cite{twoloop}. To find out  numerical values of these corrections,
one has to determine the masses of all superpartners.

Within the Constrained MSSM, imposing various constraints, one can
define the allowed region in the parameter space and calculate the
spectrum of superpartners and, hence, the radiative corrections to
the Higgs boson mass. The Higgs mass depends mainly on the following
parameters: the top mass, the squark masses, the mixing in the stop
sector, the pseudoscalar Higgs mass and $\tan\beta$. The maximum of
the Higgs mass is obtained for large $\tan\beta$, for the maximal
value of the top and squark masses and the minimal value of the stop
mixing.

\begin{figure}[htb]
\epsfig{file=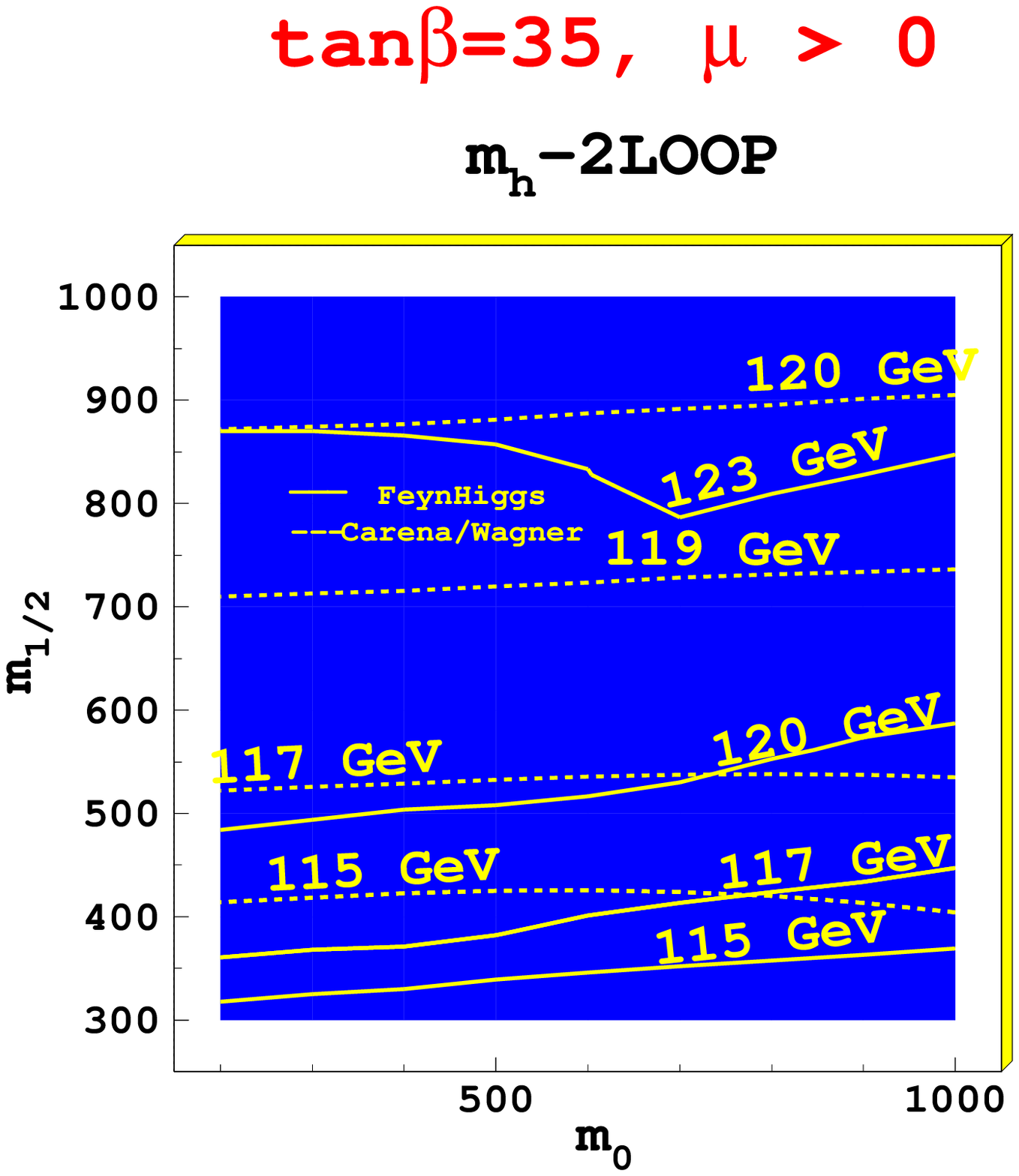,width=.49\textwidth}
\epsfig{file=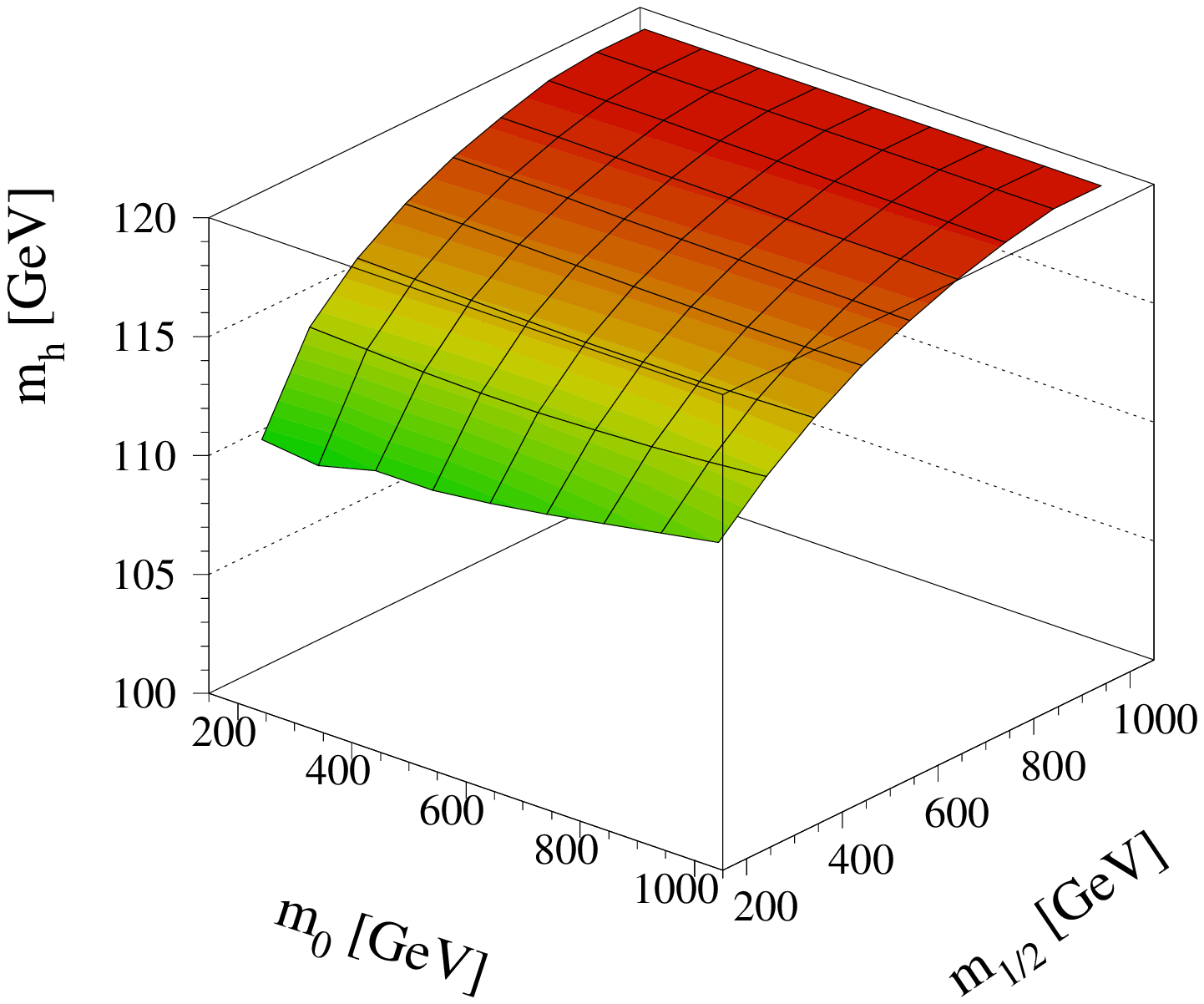,width=.49\textwidth}
\caption[]{\label{hi35} The value of the Higgs mass $m_0,m_{1/2}$ plane for the high tan
$\beta$ solution $\tan\beta=35$.}
\end{figure}

We present the value of the lightest Higgs mass in the whole $m_0,m_{1/2}$ plane for the
high $\tan\beta$ solutions in Fig.\ref{hi35}~\cite{BHGK}. One can see that it is
practically constant in the whole plane and is saturated for high values of $m_0$ and
$m_{1/2}$.

\begin{figure}[htb]
\begin{center}
 \leavevmode
  \epsfxsize=8cm \epsfysize=6cm
 \epsffile{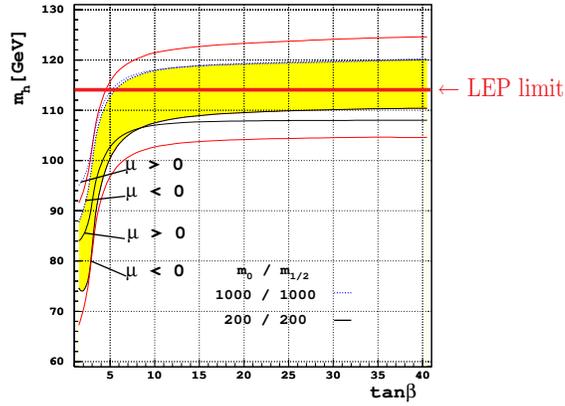}
\end{center}\vspace{-1cm}
\caption{The mass of the lightest Higgs boson as a function of $\tan\beta$}
\label{fig:mhtb}
\end{figure}

The lightest Higgs boson mass $m_h$ is shown as a function of $\tan\beta$ in
Fig.~\ref{fig:mhtb} \cite{BHGK}. The shaded band corresponds to the uncertainty from the
stop mass and stop mixing for $m_t=175$ GeV. The upper and lower lines correspond to
$m_t$=170 and 180 GeV, respectively.

%

The parameters used for the calculation of the upper limit are: $m_t=180$ GeV,
$A_0=-3m_0$ and $m_0=m_{1/2}=1000$ GeV. The lowest line of the same figure gives the
minimal values  of $m_h$. For high $\tan\beta$  the values of $m_h$ range from 105 GeV
125 GeV. At present, there is no preference for any of the values in this range but it
can be seen that the 95\% C.L. lower limit on the Higgs mass~\cite{EWWG} of 113.3 GeV
excludes $\tan\beta<3.3$.

%
%
So combining all the uncertainties discussed before the results  for
the  Higgs mass in the CMSSM can be summarized as follows:
\begin{itemize}
 \item The low $\tan\beta$ scenario ($\tan\beta <3.3$) of the
CMSSM is excluded by the lower limit on the Higgs mass of  113.3 GeV.
\item For the high $\tan\beta$ scenario the  Higgs mass is found to be in
the range from 110 to 120 GeV for $m_t=175$ GeV.
 The  central value  is found to be~\cite{BHGK}:
 $$m_h=115\pm3~ ({\rm stop\ mass})~\pm1.5~({\rm stop\
mixing})~\pm2~({\rm theory})~\pm5~({\rm top\ mass})~\rm GeV, $$
 where the errors  are the estimated standard deviations.
This prediction is independent of $\tan\beta$ for $\tan\beta
>20$ and decreases for lower $\tan\beta$.
 \end{itemize}

\begin{figure}[h]
\begin{center}
\leavevmode \epsfxsize=5.1cm 
\hspace*{-2cm} \epsffile{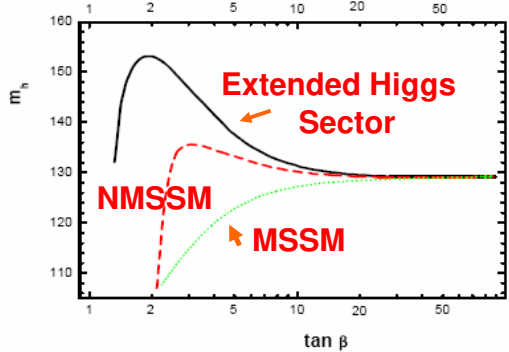}
\caption[]{ Dependence of the upper bound on the lightest Higgs boson
mass on $\tan\beta$ in MSSM (lower curve), NMSSM (middle curve) and extended SSM (upper
curve)\label{f3} }
\end{center}
\end{figure}

 However, these SUSY limits on the Higgs mass may not be so restricting if
non-minimal SUSY models are considered. In a SUSY model extended by a singlet, the
so-called Next-to-Minimal model, eq.(\ref{bound}) is modified and at the tree level the
upper bound looks like~\cite{pomarol}
\begin{equation}
  m_h^2 \simeq M_Z^2\cos^2 2\beta+ \lambda^2v^2\sin^2 2\beta,
\end{equation}
where $\lambda$ is an additional singlet Yukawa coupling. This coupling being unknown
brings us back to the SM situation, though its influence is reduced by $\sin 2\beta$. As
a result, for low $\tan\beta$ the upper bound on the Higgs mass is slightly modified (see
Fig.\ref{f3}).

Even more dramatic changes are possible in models containing non-standard fields at
intermediate scales. These fields appear in scenarios with gauge mediated supersymmetry
breaking. In this case, the upper bound on the Higgs mass may increase up to 155
GeV~\cite{pomarol} (the upper curve in Fig.\ref{f3}), though it is not necessarily
saturated.  One should notice, however, that these more sophisticated models do not
change the generic feature of SUSY theories, the presence of the light Higgs boson.

%

\section{Observation of the Higgs boson at LHC}

\begin{figure}[h]
\begin{center}
 \leavevmode
  \epsfxsize=4.5cm
 \epsffile{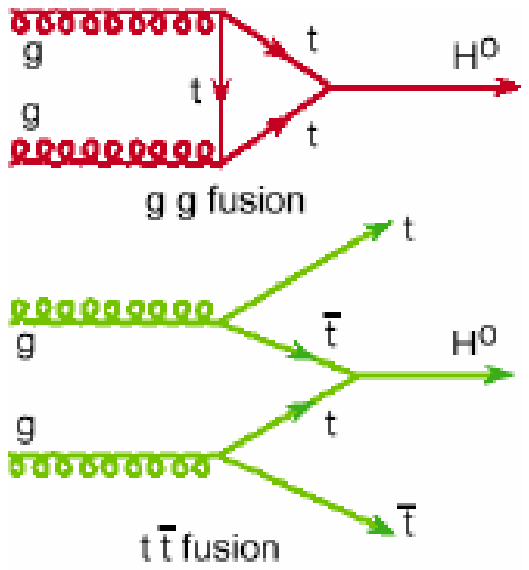}\hspace*{1cm}
 \epsfxsize=4.5cm \epsffile{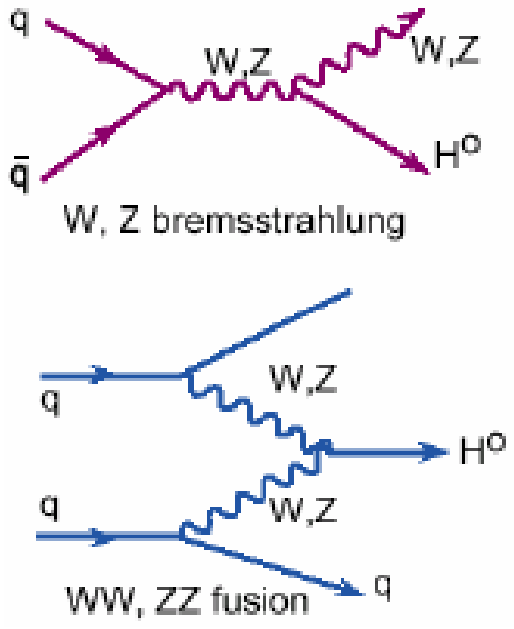}
\end{center}\vspace{-0.5cm}
\caption{The main Higgs boson production processes at hadron colliders} \label{cr}
\end{figure}

\begin{figure}[h]
\begin{center}
 \leavevmode
  \epsfxsize=9.5cm
 \epsffile{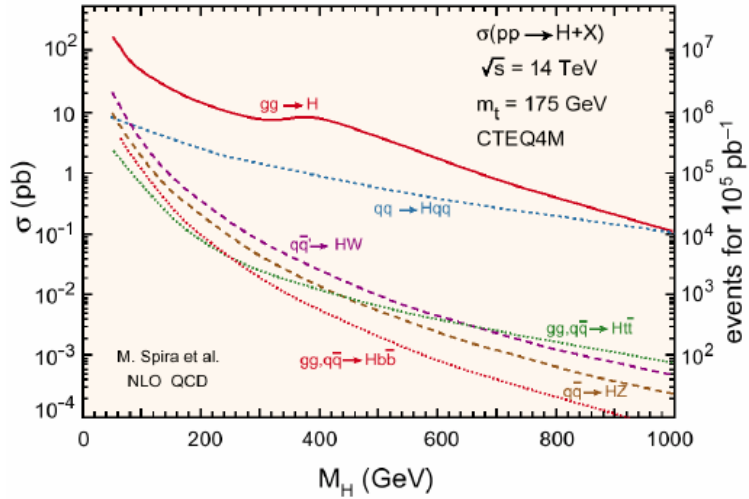}
\end{center}\vspace{-0.5cm}
\caption{The cross sections of various Higgs boson production processes at LHC}
\label{xsec}
\end{figure}

In principle, LHC will be able to cover the whole interval of SUSY
and Higgs masses up to a few TeV. However, due to severe background,
specially for the Higgs mass around the $Z$-boson mass  one needs
large integrated luminosity. From the point of view of observation
there is no much difference between the SM Higgs boson and the
lightest Higgs of the MSSM. The main production processes at hadron
colliders are shown in Fig.~\ref{cr}.
The cross section and the role of different channels depend on the mass of the Higgs
boson as shown in Fig.\ref{xsec}~\cite{spira}.

Being created the Higgs boson will decay.  The signatures of the
Higgs boson are related to the dominant decay modes which again
depend on the mass of the Higgs boson. The main decay channels are
listed below~\cite{sphicas}
$$\begin{array}{l}
\left. \begin{array}{l}
 \bullet\ H,h \to \gamma \gamma, b\bar b \ (H \to b\bar b\ \mbox{in}\ WH, t\bar t H)\\
 \bullet\ h \to \gamma \gamma\ \mbox{in}\ WH, t\bar t h \to \ell \gamma \gamma \\
 \bullet\ h,H \to ZZ^*,ZZ \to 4\ell \\
 \end{array}  \right\} \mbox{very important and promising}\\
\left. \begin{array}{ll}
 \bullet\ h,H,A -> \tau^+ \tau^- & \to (e/\mu)^+ +H^- + E_T^{miss}\\
                      &  \to e^+ + \mu^- + E_T^{miss}\\
                      &  \to H^+ +H^-+ E_T^{miss}
                        \end{array} \right\}
\mbox{inclusively in}\ b\bar b H_{SUSY}\\
 \begin{array}{l}
\bullet\ H^+ \to \tau^+\nu\ \mbox{from}\ t\bar t \\
 \bullet\ H^+ \to \tau^+\nu\ \mbox{and}\  H^+ \to t\bar b \ \mbox{for}\  M_H>M_{top}\\
 \bullet\ A \to Zh \ \mbox{with}\ h -> b\bar b; A \to \gamma\gamma
 \end{array}\end{array}$$
$$\begin{array}{l}\left. \begin{array}{l}
 \bullet\ H,A \to \tilde{\chi}_2^0\tilde{\chi}_2^0, \tilde\chi_1^+\tilde\chi_1^- \\
 \bullet\ H^+ \to  \tilde{\chi}_2^+\tilde\chi_2^0
 \end{array}\right\}\mbox{promising}\\
 \bullet\ H \to \tau^+ \tau^-\ \mbox{in}\ WH, t\bar t H
\end{array} $$

\begin{figure}[h]
\begin{center}
 \leavevmode
 \epsfxsize=7.5cm \epsffile{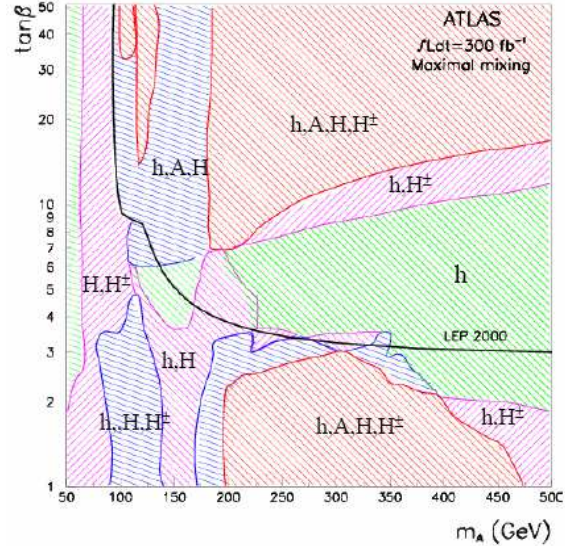}
\end{center}\vspace{-0.5cm}
\caption{Exclusion plots for LHC hadron collider for different Higgs decay modes}
\label{lhc}
\end{figure}

The LHC will either discover the SM or the MSSM Higgs boson, or
prove their absence. In terms of exclusion plots shown in
Fig.~\ref{lhc} the LHC collider will cover the whole region of SUSY
parameter space~\cite{LHC}. Various decay modes allow one to probe
different areas, as shown in Fig.~\ref{lhc}, though the background
will be very essential.

\section{Conclusion}

The LHC hadron collider will have all the possibilities for
important discoveries already in the first year of its operation
(one day of LHC with the luminosity of
$10^{33}\mbox{cm}^{-2}\mbox{s}^{-1}$ is equivalent to 10 years of
work of the previous accelerator). Supersymmetry, if the scenaria
described above are realized, might be discovered almost
immediately. Slightly more complicated is the situation with the
Higgs boson. Therefore the stable functioning of the accelerator
with high luminosity is crucial. However, to get the desired result
one need enormous efforts on data processing and calculation of the
background processes within the Standard Model at the center of mass
energy of 14 TeV.

\bigskip \bigskip
Financial support from Russian Foundation of Basic Research (grant
05--02--17603) and the grant of the President of Russian Federation
for support of leading scientific schools (NSh--5362.2006.2) is
kindly acknowledged. We are grateful to the organizers of the ITEP
Winter school for fruitful atmosphere and pleasant conditions.

\end{document}